\documentclass{article}
\usepackage{graphicx}
\usepackage{dcolumn}
\usepackage{bm}
\usepackage{epsfig}
\usepackage{slashed}
\usepackage{subfigure}
\usepackage{epsfig}
\usepackage{slashed}
\usepackage{amssymb}
\usepackage{multirow}
\usepackage{tabularx}
\usepackage{amsmath}

\begin{document}
\title{ $\boldsymbol{W\gamma}$ and $\boldsymbol{Z\gamma}$ production and limits on anomalous $\boldsymbol{WW\gamma}$, $\boldsymbol{ZZ\gamma}$ and $\boldsymbol{Z\gamma\gamma}$ couplings with D0 detector}
\author{ 
\large{Xuebing Bu}\\
\small{Fermi National Accelerator Laboratory, Batavia, IIlinois 60510, USA}\\
\small{xbbu@fnal.gov}
}
\date{May 8, 2012}
\maketitle
The recent D0 results on $W\gamma$ and $Z\gamma$ production are presented.
First, the cross section and the difference in rapidities
between photons and charged leptons for inclusive $W(\to l\nu)+\gamma$ 
production in $e\gamma$ and $\mu\gamma$ final states are discussed,
then are the cross section and differential cross section as a function of
photon transverse momentum for $Z\gamma \to l^{+}l^{-} (l=e,\mu)$
process. Finally, I present the limits on anomalous $WW\gamma$, $ZZ\gamma$
and $Z\gamma\gamma$ couplings.

\section{Introduction}
The electroweak component of standard model (SM) has been remarkably successful
in describing experimental results.
The self-interaction of the gauge bosons (the $W$, the $Z$, and the photon)
is a consequence of the non-Abelian $SU(2)_{L} \times U(1)_{Y}$ gauge symmetry
of the SM. The gauge boson self-interactions appear as vertices involving
three gauge bosons, and result in the production of pairs of bosons.
The $WW\gamma$, $ZZ\gamma$ and $Z\gamma\gamma$ vertices are examples for that
self-interactions of gauge bosons, and very sensitive to new physics.
For instance, we could use the process
$p\bar{p} \to W\gamma \to l\nu\gamma$ $(l=e,\mu)$
to study the $WW\gamma$ vertex, and search for any anomalous departure from SM
$WW\gamma$ coupling.
In particular, because the Z boson carries no electric charge, a coupling
between a $Z$ boson and a photon is not permitted in SM.

In this review we summarize recent D0 results in measurements of
cross section and the difference in rapidities between photons and leptons
for inclusive $W(\to l\nu)+\gamma$ $(l=e,\mu)$ production,
and the cross section and differential cross section
as a function of photon momentum for $Z\gamma \to l^{+}l^{-} (l=e,\mu)$
production, as well as the limits on anomalous $WW\gamma$,
$ZZ\gamma$ and $Z\gamma\gamma$ couplings \cite{wg_prl,zg_prd}.

\section{The D0 detector}
\begin{figure}[h]
  \begin{center}
    \epsfig{file=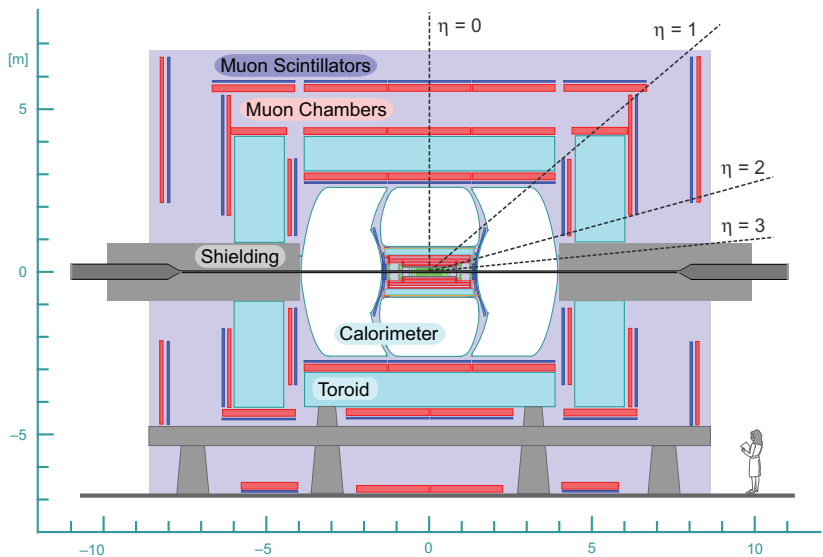,scale=0.8}
  \end{center}
  \caption{\small (color online). The D0 Detector.}
  \label{fg:d0_detector}
\end{figure}

The D0 detector is a multi-purpose particle detector. It has
been constructed to study proton-antiproton collisions at a center
of mass energy $\sqrt{s} = 1.96$ TeV.
The D0 detector \cite{d0det} comprises a central tracking
system in a 2~T superconducting solenoidal magnet,
surrounded by a central preshower (CPS) detector, a
liquid--argon sampling calorimeter, and an outer muon system.
Fig. \ref{fg:d0_detector} is the overview of D0 detector.
The tracking system, a silicon microstrip tracker (SMT)
and a scintillating fiber tracker (CFT), provides coverage
for charged particles in the pseudorapidity range
$|\eta| < 3$~\cite{d0_coordinate}.
The CPS is located immediately before the inner layer of the calorimeter,
and has about one radiation length of absorber, followed by several
layers of scintillating strips.
The calorimeter consists of a central sector (CC) with
coverage of $|\eta|<1.1$, and two end calorimeters (EC)
covering up to $|\eta| \approx 4.2$.
The electromagnetic (EM) section of the
calorimeter is segmented into four longitudinal layers
(EM$i$, $i=1,4$) with transverse
segmentation of
$\Delta\eta\times\Delta\phi = 0.1\times 0.1$~\cite{d0_coordinate},
except in EM3,
where it is $0.05\times 0.05$.
The muon system resides beyond the calorimeter and consists of a
layer of tracking detectors and scintillation trigger counters
before 1.8 T iron toroidal magnet, followed by two
similar layers after the toroid. The coverage of the muon
system corresponds to $|\eta| < 2$.
Luminosity is measured using plastic scintillator arrays located
in front of the
EC cryostats, covering $2.7<|\eta|<4.4$.
The data acquisition system
consists of a three-level trigger, designed to accommodate the high
instantaneous luminosity.

\section{$W\gamma$ production}
\label{sec:wg}
\begin{figure}[htbp]
  \begin{center}
  \epsfig{file=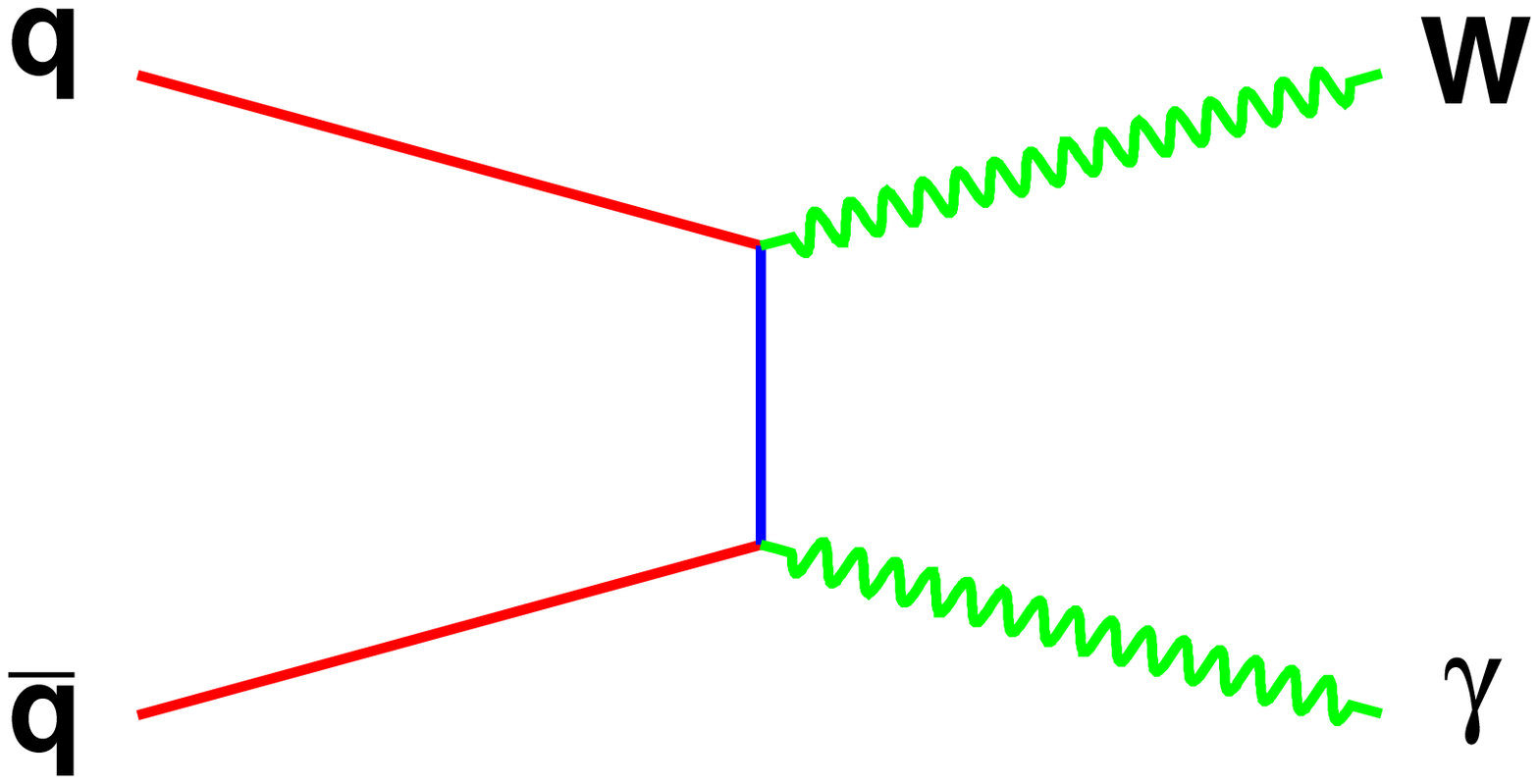,scale=0.3}
  \epsfig{file=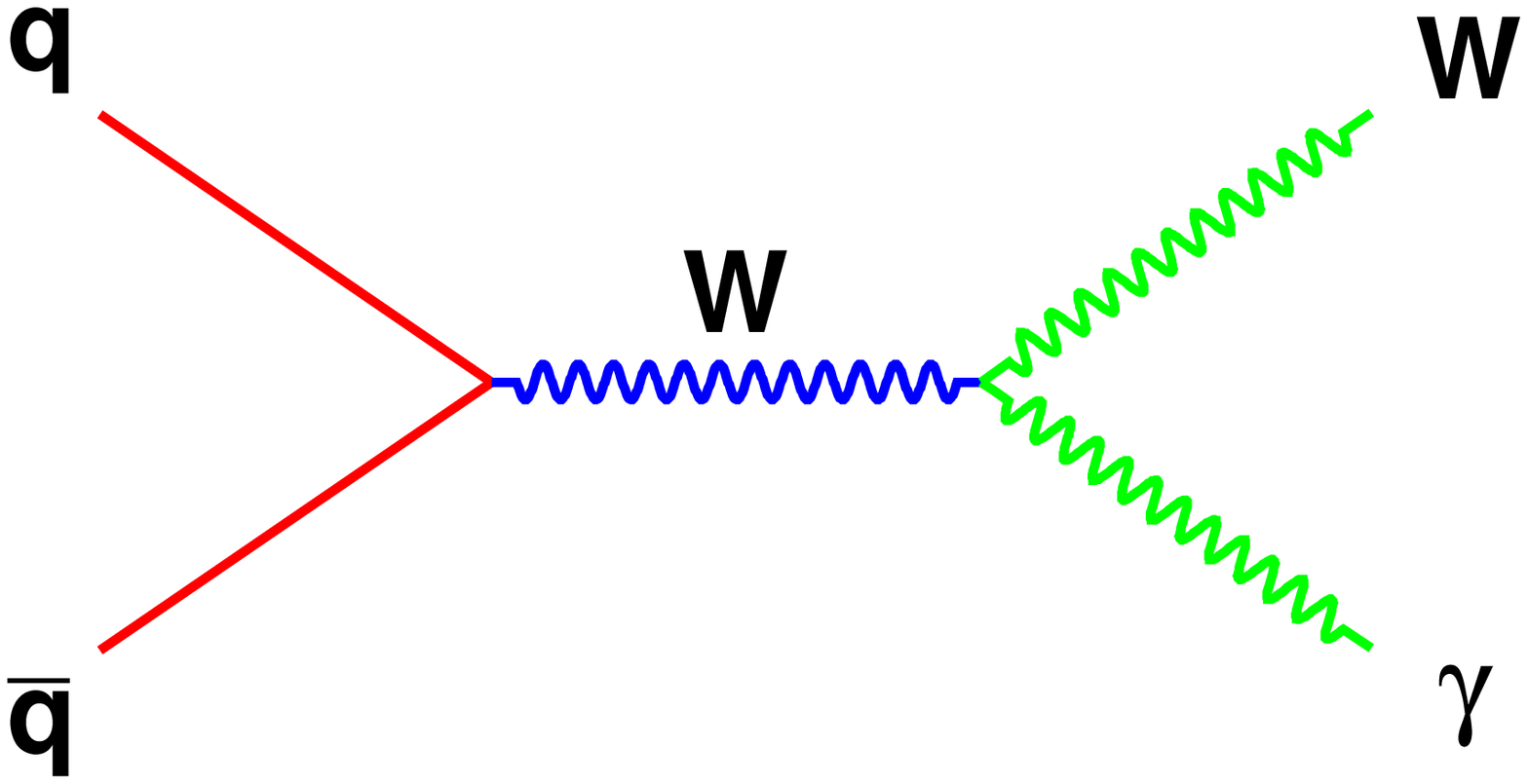,scale=0.3}
  \epsfig{file=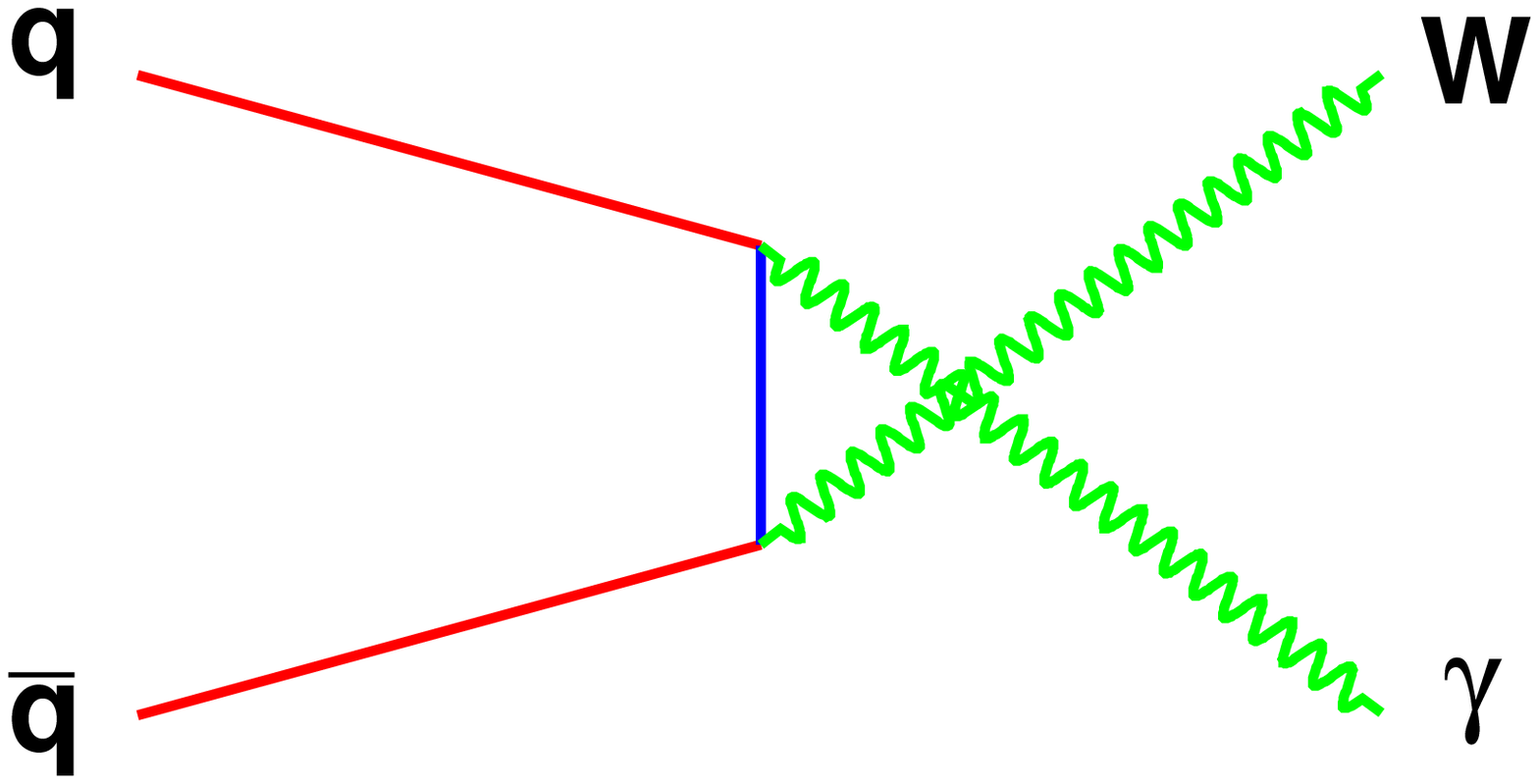,scale=0.3} 
  \end{center}
   \caption{\small (color online).
Feynman diagrams for prompt $W\gamma$ production.
}
  \label{figure-wg}
\end{figure}

\begin{figure}[htbp]
  \begin{center}
  \epsfig{file=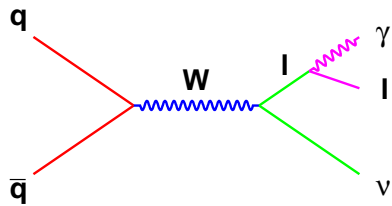,scale=0.3}
  \end{center}
   \caption{\small (color online).
Feynman diagram for $W\gamma$ production via final state radiation.
}
  \label{figure-wg-fsr}
\end{figure}

\begin{figure}[htbp]
\begin{center}
\epsfig{file=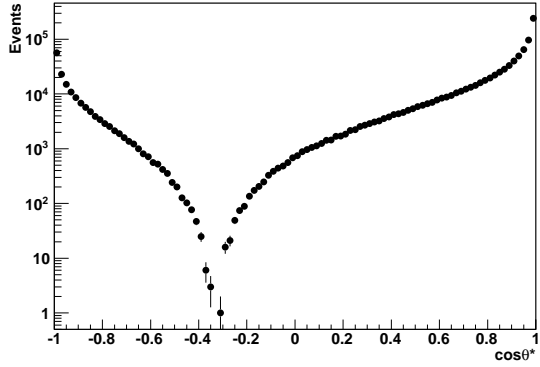,scale=0.4}
\end{center}
\caption{\small
$cos\theta^{*}$ between the W and incoming quark
in W$\gamma$ rest frame from {\sc pythia} \cite{pythia-ref}.}
\label{WG_RAZ1}
\end{figure}

\begin{figure}[htbp]
\begin{center}
\epsfig{file=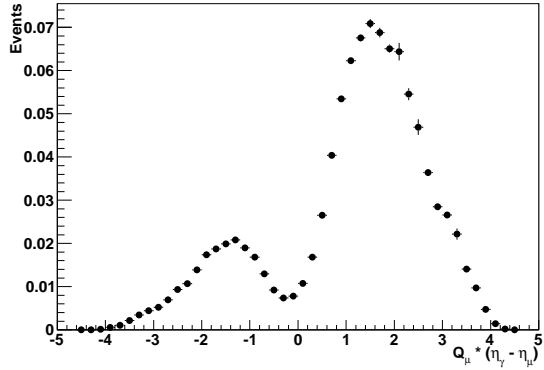,scale=0.4}
\end{center}
\caption{\small
Charge-signed photon-lepton rapidity difference
from next-to-leading order W$\gamma$ generator \cite{baur-ref1, baur-ref2} }
\label{WG_RAZ2}
\end{figure}

In the SM, tree level production of a photon in association
with a $W$ boson occurs due to
prompt $W\gamma$ production via the diagrams shown in Fig.~\ref{figure-wg} or
via final state radiation (FSR), where a lepton from the $W$ boson decay
radiates a photon (shown in Fig. \ref{figure-wg-fsr}).
At leading order (LO) of SM prediction,
the interference between the amplitudes
in Fig.~\ref{figure-wg} produces a zero in the total $W\gamma$ yield
at a specific angle $\theta^{*}$ between the $W$ boson
and the incoming quark \cite{raz-ref} in the $W\gamma$ rest frame
(see Fig. \ref{WG_RAZ1}).
Since in hadronic collisions the longitudinal momenta of neutrinos from $W$
decay cannot be measured, the angle $\theta^{*}$ at which the radiation
amplitude is zero
is difficult to measure directly.
However, the radiation amplitude zero (RAZ) is also visible in
the charge-signed
photon-lepton rapidity difference as a dip around $-$1/3 \cite{baur-ref1}
(see Fig. \ref{WG_RAZ2}).

In this review, I present the measurements of the cross section and
the distribution of the charge-signed rapidities difference between
photon and lepton using data corresponding to an integrated luminosity
of 4.2 fb$^{-1}$ collected by D0 detector at the Fermilab Tevatron Collider.

\subsection{Event selection}
For the electron channel with the $W$ boson decaying into an electron and
a neutrino, a suite of single-electron triggers are used to collect the
candidate events.
The electrons are selected by requiring an EM cluster in either
the CC ($|\eta| < 1.1$) or EC ($1.5 < |\eta| <2.5$)
with transverse momentum $p_T > 25$~GeV.
Electron candidates are required to
be isolated in both the calorimeter and the tracker,
have a shower shape consistent with that
of an electron, and a spatial match to a track.
Further a multivariate likelihood discriminant and an artificial
neural network are used to further reject background from jets
misidentified as electrons.
To suppress the $Z/\gamma^{*} \to ee$ background,
the event missing transverse energy, $\slashed E_{T}$,
must exceed 25 GeV,
the transverse mass of the $W$ boson, $M_T$,
must exceed 50 GeV and
the azimuthal angle between the electron and photon is required to be
$\Delta\phi_{e\gamma}<2$.

For the muon channel with the $W$ boson decaying into an muon and a neutrino,
a suite of single-muon triggers are used to collect the candidate events.
The muons are required to be within $|\eta| < 1.6$,
isolated in both the tracker and the calorimeter,
and matched to a track with transverse momentum $p_T > 20$ GeV.
To suppress the $Z/\gamma^{*} \to \mu\mu$ background,
the $\slashed E_{T}$ in the event must exceed 20 GeV,
$M_T$ must exceed 40 GeV, and there must be no additional muons
or tracks with $p_T > 15$ GeV.

The photon candidates in both the electron and muon channels are
required to have transverse momentum $p_T^{\gamma} > 15$ GeV.
In addition, photon candidates are required to be either in the
CC~($|\eta| < 1.1$) or EC~($1.5 < |\eta| <2.5$),
and be isolated in both the calorimeter and the tracker.
Furthermore, the output of an artificial neural network
($O_{NN}$)~\cite{HggPRL},
that combines information from a set of variables sensitive to
differences between photons and jets
in the tracking detector, the calorimeter, and the CPS detector,
is required to be larger than 0.75.

To suppress background from FSR, the photon and the lepton
must be separated by $\Delta R_{l\gamma} > 0.7$,
and the three-body transverse mass (see Eq. \ref{eq:M3})
of the photon, lepton, and
missing transverse energy must exceed 110 GeV.

\begin{equation}
\label{eq:M3}
M_{T}^{l\gamma\slashed E_{T}} = \sqrt{ ( \sqrt{M^{2}_{l\gamma} + 
                                 |\bf{ p_{T}(\gamma) + p_{T}(l) }|^{2}} 
                                 + \slashed E_{T})^{2}
                                 - |\bf{ p_{T}(\gamma) + p_{T}(l) } + \slashed E_{T}|^{2} }
\end{equation}

\subsection{Backgrounds}
There are four major sources of background in this analysis:
(i) events with $el + X$ final state,
where the electron is misidentified as photon due to tracking inefficiency;
(ii) $W$+jet production, where the jet is misidentified as the photon;
(iii) $Z_{ll}\gamma$ production, where one of the lepton is lost;
(iv) $W_{\tau\nu}\gamma$ production, where $\tau$ further decays
to $e$ or $\mu$.

The $el + X$ background is composed of events
where the electron is misidentified as
a photon due to tracking inefficiencies,
and mainly comes from the di-boson production.
To estimate its contribution, an orthogonal data sample is selected
by requiring the EM cluster be matched with a good track.
Then the ratio for EM cluster matching a good track and passing the
photon no-track requirement is measured from the $Z \to ee$ data with
parameterizing as a function of $\eta$.
Finally, the $el + X$ contribution is calculated with multiplying the ratio
on the orthogonal sample.

The dominant background for this analysis is W+jet production.
Two different data driven methods have been used to estimate the contribution.
In method one, an orthogonal data sample (W+bad photon) is selected by
reversing the photon track isolation or shower width requirement.
Then the ratio of jet passing the good photon selection criteria
and failing the track isolation or shower width requirement
is measured from the di-jet data.
This ratio is measured as a function of $p_{T}$
in 5 $\eta$ regions for photon in CC and EC respectively.
The final W+jet contribution is obtained with applying these ratios
to the selected W+bad photon data events.
In method two, a fit is performed on the photon $O_{NN}$ distributions
in 5 $\eta$ regions for photon in CC and EC respectively.
The photon $O_{NN}$ templates are obtained
from photon and jet Monte Carlo (MC) simulation, since the $O_{NN}$ is well
modelled \cite{HggPRL, XggPLB}.
The results from these two methods are consistent,
considering the poor statistics of template fitting method,
the results from reversing photon quality cuts is used as the default.

Small backgrounds from $Z_{ll}+\gamma$, where one of the leptons from Z decays
is lost, and $W_{\tau\nu}+\gamma$, where the $\tau$ decays to a $e$ or $\mu$,
are estimated from {\sc pythia} Drell-Yan $Z/\gamma* \to \mu\mu$ and
$W_{\tau\nu}+\gamma$ MC respectively.

\subsection{Signal}
The signal is simulated using the
Baur and Berger LO event generator~\cite{baur-ref3},
interfaced to {\sc pythia}~\cite{pythia-ref}
for subsequent parton showering and hadronization.
To avoid the double counting of the FSR events, the diagram
corresponding to FSR of photons is disabled in {\sc pythia}.
The shape and normalization of the signal $p_T^{\gamma}$ spectrum
are reweighted to the next-to-leading order (NLO) prediction~\cite{baur-ref2}.
The acceptance of the kinematic and geometric requirements
for this analysis is calculated using
this $p_T^{\gamma}$-weighted signal MC.
All MC events are generated using the CTEQ6L1 \cite{cteq-ref}
parton distribution functions (PDF),
followed by a {\sc geant} \cite{geant-ref}
simulation of the D0 detector.

\subsection{Systematic uncertainties and results}
The dominant systematic uncertainties considered in this analysis are:
\begin{itemize}
 \item 6.1\% uncertainty on the total luminosity;
 \item 5\% uncertainty on the single lepton trigger efficiency;
 \item 3\% uncertainty on the photon identification;
 \item 3\% uncertainty on the lepton identification;
 \item 0.9\% uncertainty on the track veto;
 \item $\sim 10$ \% uncertainty on estimation of W+jet background.
\end{itemize}

The number of predicted and observed events in both the electron
and muon channels are summarized in
Table \ref{table:Nevents}.

\begin{table}[t]
\begin{center}
\begin{tabular}{ccc}
    \hline \hline
    & $e\nu\gamma$ channel& $\mu\nu\gamma$ channel \\
    \hline
    $W+$jet& 33.9 $\pm$ 3.7 & 64.6 $\pm$ 6.8 \\
    $leX$ & 1.1 $\pm$ 0.6 & 2.1 $\pm$ 0.7 \\
    $Z\gamma \to ll\gamma$& 1.8 $\pm$ 0.3 & 17.6 $\pm$ 1.9 \\
    $W\gamma \to \tau\nu\gamma$& 2.3 $\pm$ 0.3 & 5.4 $\pm$ 0.6 \\
   \hline
   Total background& 39.1 $\pm$ 3.8 &89.7 $\pm$ 7.2 \\
   SM $W\gamma$ prediction& 150.9 $\pm$ 13.8 & 282.1 $\pm$ 25.4 \\
   Data& 196 & 363 \\
    \hline \hline
  \end{tabular}
  \caption{ \small
   Number of predicted and observed events with
statistical and systematic uncertainties.
(Table from Ref. \cite{wg_prl}, see text)
}
  \label{table:Nevents}
\end{center}
\end{table}

The cross section multiplied by the branching fraction for
the process $p\bar{p} \to W\gamma + X \to l\nu\gamma +X$ is measured to be
7.6~$\pm$~0.4~(stat.)~$\pm$~0.6~(syst.)~pb,
which is in good agreement with the SM expectation of 7.6~$\pm$~0.2~pb
for $E_{T}^{\gamma} >$ 15 GeV and $\Delta R_{l\gamma} >$ 0.7.

\begin{figure}[htbp]
 \centering
\hspace*{-9mm} \includegraphics[scale=0.5]{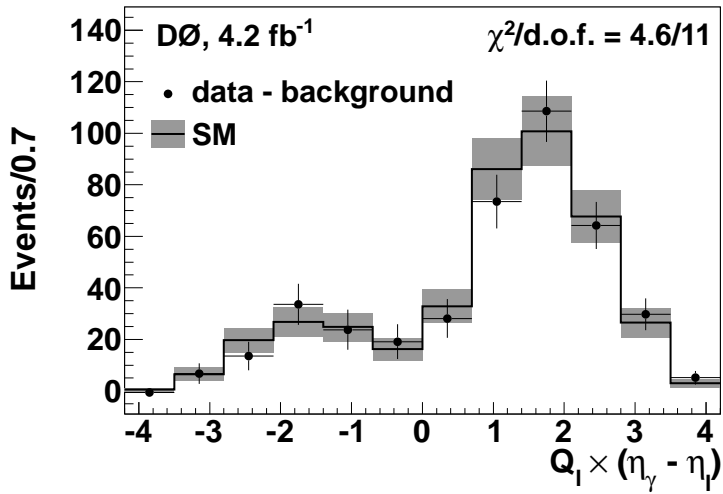}
~\\[-4mm]
   \caption{\small
   The charge-signed photon-lepton rapidity difference
($Q_{l} \times (\eta_{\gamma} - \eta_{l})$, where
$Q_{l}$ is the charge of the lepton)
in background-subtracted data compared to the SM expectation for
the combined electron and muon channels.
The background-subtracted data are shown as black points
with error bars representing their total uncertainties.
The SM signal prediction is given by the solid line,
with the shaded area representing its uncertainty.
(Figure from Ref. \cite{wg_prl}, see text)
}
  \label{figure-raz}
\end{figure}

The charge-signed photon-lepton rapidity difference
for the combined electron and muon channels is shown in Fig.~\ref{figure-raz}.
The events with EC electrons are excluded from Fig.~\ref{figure-raz}
due to the significant charge mis-identification.
The background-subtracted data are in good agreement with the SM prediction,
and a $\chi^{2}$ test comparing the background-subtracted data with the
SM prediction
yields 4.6 for 11 degrees of freedom.

\section{$Z\gamma$ production}
\begin{figure}[htbp]
\centering
\subfigure[]{\label{fig:ISR1}\includegraphics[width=0.35\textwidth]{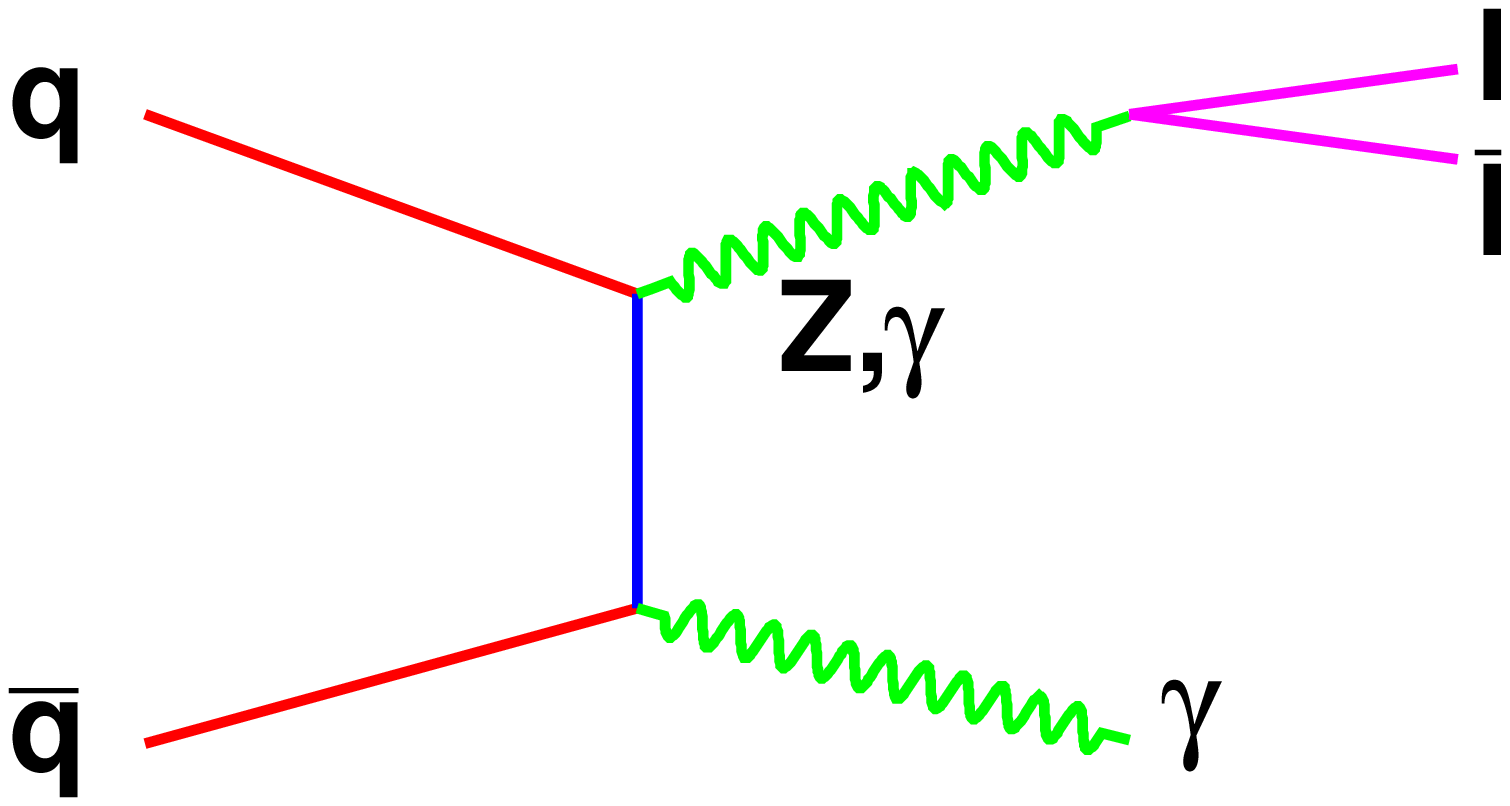}}
\subfigure[]{\label{fig:ISR2}\includegraphics[width=0.35\textwidth]{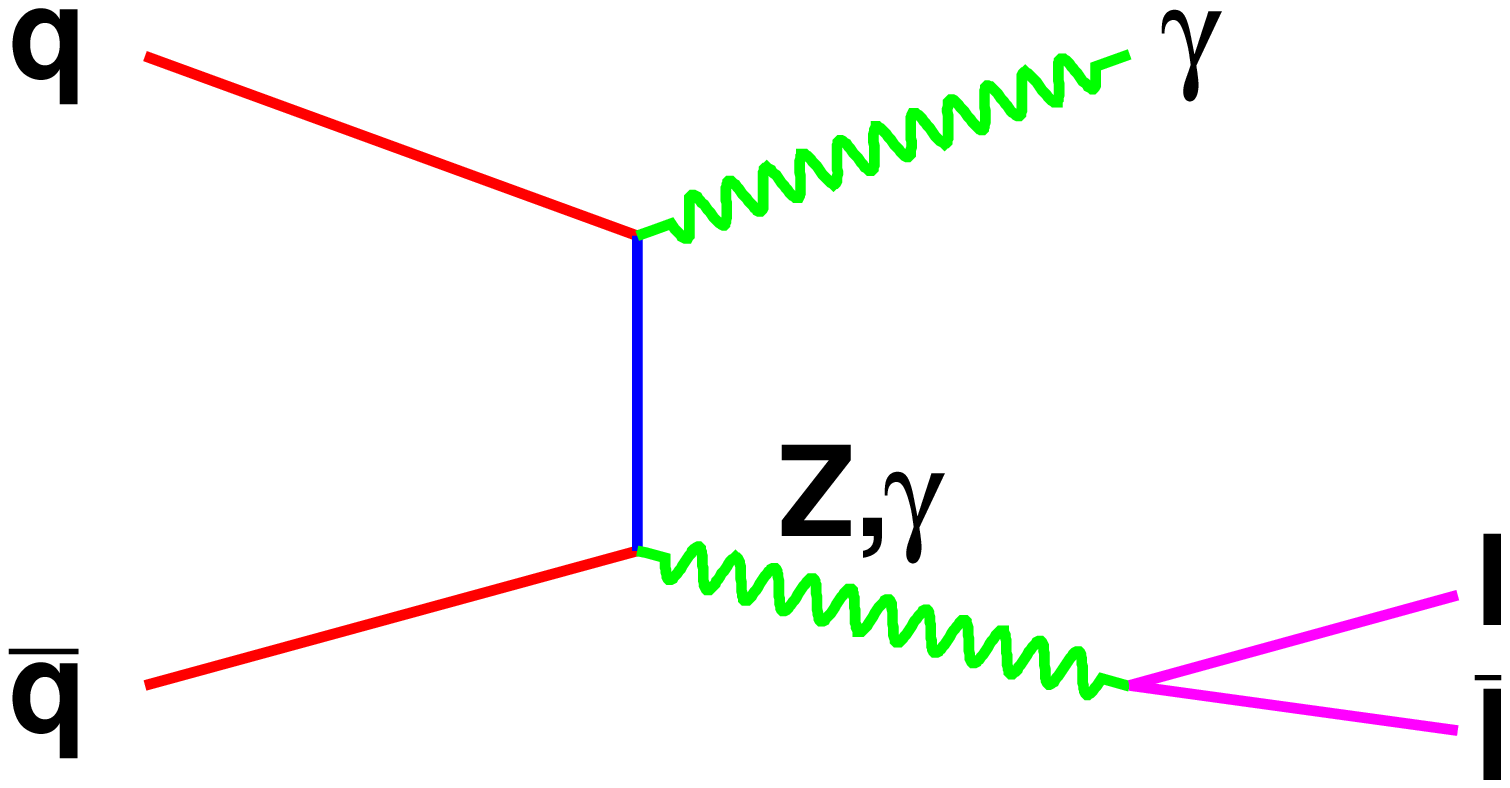}}
\subfigure[]{\label{fig:FSR1}\includegraphics[width=0.35\textwidth]{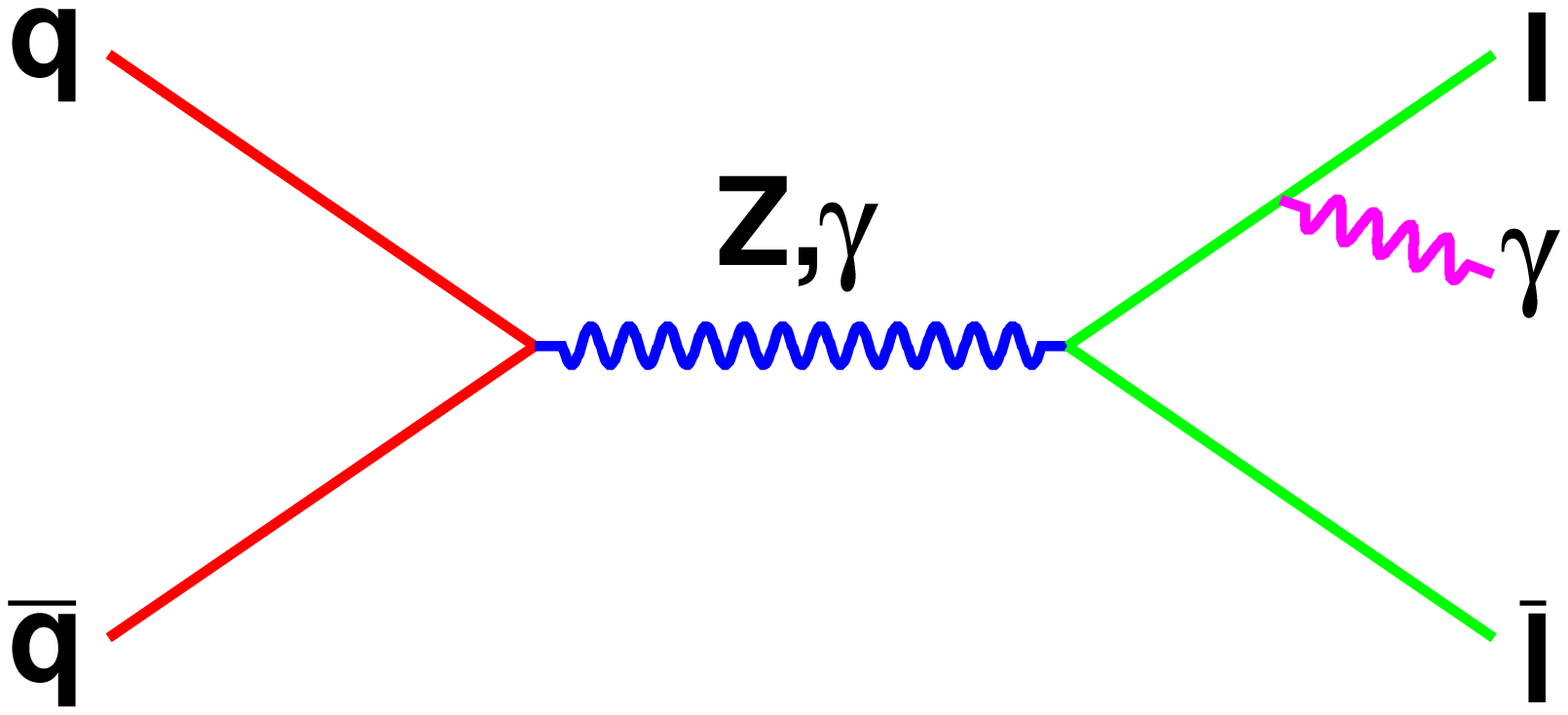}}
\subfigure[]{\label{fig:FSR2}\includegraphics[width=0.35\textwidth]{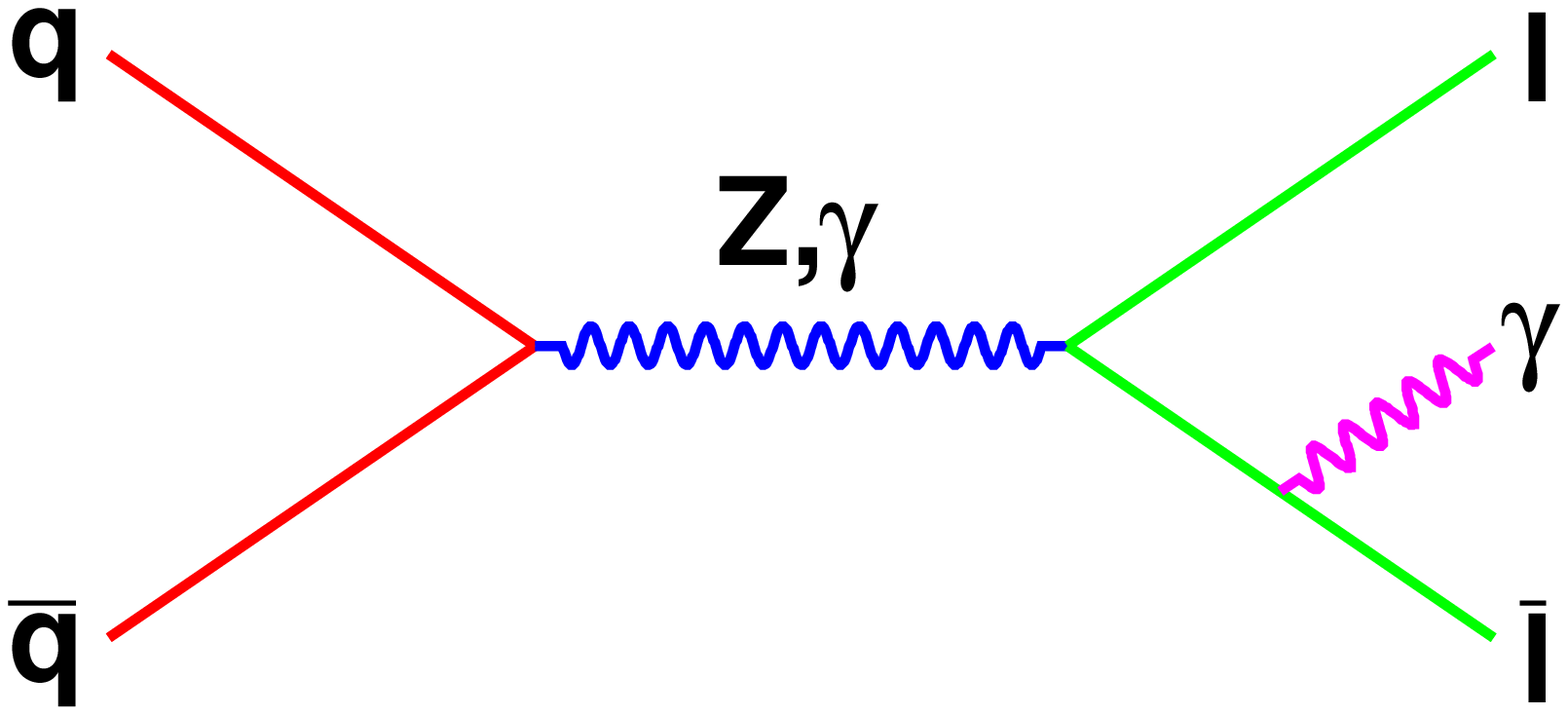}}
\caption{\small
  Feynman diagrams for the tree-level $Z(\gamma) \to l^{+}l^{-}\gamma$
  processes in SM:
  (a) and (b) describe the initial state radiation,
  (c) and (d) describe the final state radiation.
 }
\label{fig:ZG_SM}
\end{figure}

In the SM, production of a photon in association with a $Z$ boson
occurs due to radiation of a photon
from an incoming quark (see Figs. \ref{fig:ISR1} and \ref{fig:ISR2}),
or from final state radiation of a
lepton of the outgoing $Z$ boson (see Figs. \ref{fig:FSR1} and \ref{fig:FSR2}).

In this review, I present the measurements of the inclusive cross section
and differential cross section for $Z\gamma$ production in the
electron and muon channels
using a data sample corresponding to an integrated luminosity
of 6.2 fb$^{-1}$ collected at
$\sqrt{s} = 1.96$ TeV by the D0 detector at the Fermilab Tevatron Collider.

\subsection{Event selection}
For electron channel with $Z$ boson decaying to two electrons, the same
single-electron triggers and electron selection criteria as $W\gamma$
analysis described in Section \ref{sec:wg} is used.
Two electron candidates with transverse momentum $p_T > 15$ GeV
are selected, and the highest $p_T$ electron must have $p_T > 25$ GeV.

For muon channel with $Z$ boson decaying to two muons, the same single-muon
triggers and muon selection criteria as $W\gamma$
analysis is used, except the $\eta$
coverage is extended to 2. Two muon candidates with transverse momentum
$p_{T} > 15$ GeV are selected, and the highest $p_{T}$ muon must have
$p_{T} > 20$ GeV.

Photon candidates in both electron and muon channels are required to
be in CC and have
transverse momentum $p_T > 10$ GeV. By comparison with $W\gamma$ analysis,
same photon selection criteria are used, except the $O_{NN}$ is
required to be greater than 0.1 instead of 0.75.

In addition, the dilepton invariant mass, $M_{ll}$, is required
to be greater than 60 GeV, and the photon must be separated from each lepton
by $\Delta R_{l\gamma} > 0.7$.
In the end, 1002 and 1000 data events are selected
in electron and muon channels respectively.
In order to reduce the contribution of FSR, subset data samples are
defined with the requirement that the reconstructed three-body invariant mass,
$M_{ll\gamma}$, exceed 110 GeV.
With this additional requirement, 304 and 308 data events are selected
in the electron
and muon channels, respectively.

\subsection{Background}
The dominant background for this analysis is the Z+jet, where the jet is
misidentified as a photon.
For electron channel, same as $W\gamma$ analysis, the Z+jet background
is estimated from an orthogonal data sample by reversing the photon
quality cuts.
For muon channel, a $2 \times 2$ matrix method is used to estimate
the Z+jet background contribution.
The Z+jet background is also estimated through a fit
to the shape of the $O_{NN}$ distributions in data
for both electron and muon channels. The results are in good
agreement with those obtained from the reversing and matrix methods.

\subsection{Total cross section}
To eliminate the uncertainties of lepton trigger efficiencies, reconstruction
efficiencies and integrated luminosity, the total cross section for
$ll\gamma$ production is obtained from the ratio of the acceptance-corrected
$ll\gamma$ rate for
$M_{ll} > 60$ GeV, $\Delta R_{ll\gamma} > 0.7$,
$p_T^\gamma > 10$ GeV/$c$, and $|\eta^\gamma| < 1$,
to the total acceptance-corrected dilepton rate for $M_{ll} > 60$ GeV.
Thus the total cross section for $Z\gamma$ is estimated using the ratio
to multiply the inclusive $Z/\gamma^{*} \to ll$ production theoretical
cross section:

\begin{equation}
\label{xsect}
\sigma_{Z\gamma} = \frac{\kappa \cdot N_{ll\gamma}^{\text{data}} \cdot (A\times\epsilon_{ID})_{ll\gamma}^{-1}} 
  {N_{ll}^{\text{data}} \cdot (A\times\epsilon_{ID})_{ll}^{-1}}  
 \times \sigma_Z
\end{equation}

Here, $N_{ll}^{\text{data}}$ and $N_{ll\gamma}^{\text{data}}$
are the number of selected $Z$ and background-subtracted $Z\gamma$
events in data sample, respectively.
The $\sigma_Z$ is the theoretical cross section for
inclusive $Z/\gamma^{*} \to ll$ production.
The factor $\kappa$ corrects for the detector resolution effects
that would cause events not to pass the selections on the generator-level
quantities, but to pass the reconstruction requirements,
where the photon energy resolution in the low $p_{T}$ is the dominant source.
The factors
$(A\times\epsilon_{ID})_{ll\gamma}$ and $(A\times\epsilon_{ID})_{ll}$
provide the fraction of events
that pass the analysis requirements, with all acceptances measured
relative to the kinematic requirements at the generator level for the
$ll\gamma$ and $ll$ final states, respectively.
The {\sc pythia} $Z/\gamma^{*}\rightarrow ll$ events are used to calculate
the $A\times\epsilon_{ID}$ after reweighting the $p_T^Z$ spectrum to the
observed one.

The total cross section has been measured with and without the
$M_{ll\gamma} > 110$ GeV requirement to reflect the FSR effect.
The corresponding results are shown in Tables \ref{xsect} and
\ref{xsect_llg110}.
The measured cross section are consistent with the
NLO {\sc mcfm} \cite{mcfm-ref}.

\begin{table}[h!]
\caption{\label{xsect} Summary of the total cross-section measurements, when no $M_{\ell\ell\gamma}$ requirement is applied, for individual channels, combined channels, and the NLO {\sc mcfm} calculation with associated PDF and scale uncertainties. (Table from Ref. \cite{zg_prd}, see text) }
\begin{center}
  \begin{tabular}{   l l l }
  \hline 
  \hline
    & & \multicolumn{1}{c}{$\sigma_{Z\gamma} \times \mathcal{B}$ [fb]} \\ \hline 
    \multicolumn{2}{l}{$ee\gamma$ data} & 1026 $\pm$ 62 (stat.) $\pm$ 60 (syst.) \\
     \multicolumn{2}{l}{$\mu\mu\gamma$ data} & 1158 $\pm$ 53 (stat.) $\pm$ 70 (syst.) \\ \hline
     \multicolumn{2}{l}{$\ell\ell\gamma$ combined data} & 1089 $\pm$ 40 (stat.) $\pm$ 65 (syst.) \\ \hline
    \multicolumn{2}{l}{NLO {\sc mcfm}} & 1096 $\pm$ 34 (PDF) $^{+2}_{-4}$ (scale) \\ \hline \hline
\end{tabular}
\end{center}
\end{table}

\begin{table}[h!]
\caption{\label{xsect_llg110} Summary of the total cross-section measurements, with the $M_{\ell\ell\gamma}>110$  GeV/$c^2$ requirement, for individual channels, combined channels, and the NLO {\sc mcfm} calculation with associated PDF and scale uncertainties. (Table from Ref. \cite{zg_prd}, see text)}
\begin{center}
  \begin{tabular}{  l l l }
  \hline 
  \hline
     & & \multicolumn{1}{c}{$\sigma_{Z\gamma} \times \mathcal{B}$ [fb]} \\ \hline
    \multicolumn{2}{l}{$ee\gamma$ data} & 281 $\pm$ 17 (stat.) $\pm$ 11 (syst.) \\
    \multicolumn{2}{l}{$\mu\mu\gamma$ data} & 306 $\pm$ 28 (stat.) $\pm$ 11 (syst.) \\ \hline
     \multicolumn{2}{l}{$\ell\ell\gamma$ combined data} & 288 $\pm$ 15 (stat.) $\pm$ 11 (syst.) \\ \hline
     \multicolumn{2}{l}{NLO {\sc mcfm}} & 294 $\pm$ 10 (PDF) $^{+1}_{-2}$ (scale) \\ \hline \hline
\end{tabular}
\end{center}
\end{table}

\subsection{Differential cross section $d\sigma/dp_{T}^{\gamma}$}
\begin{figure}[htbp]
\centering
\includegraphics[width=0.45\textwidth]{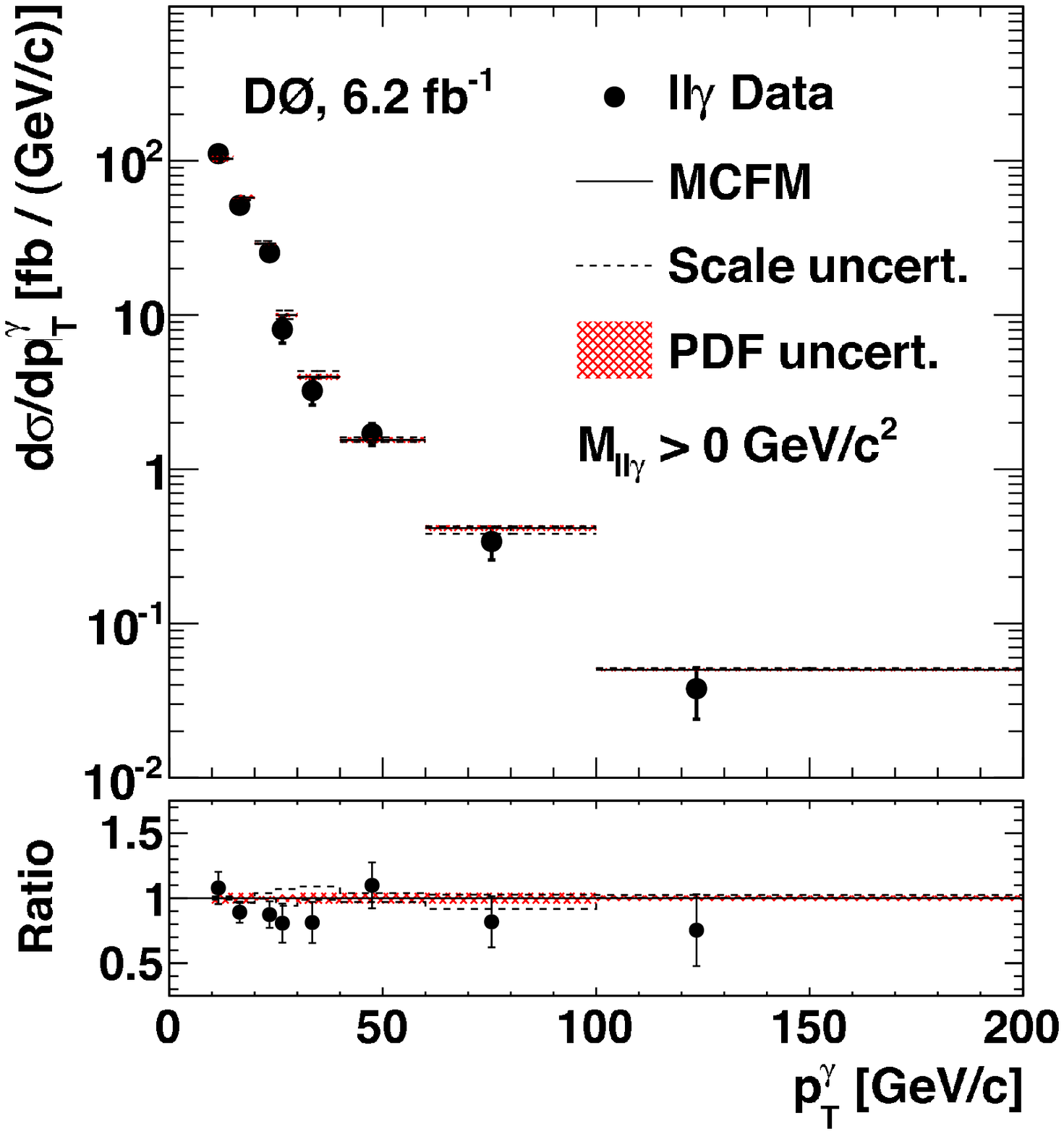}
\caption{Unfolded $d\sigma/dp_T^\gamma$ distribution with no $M_{ll\gamma}$ requirement for combined electron and muon data compared to the NLO {\sc mcfm} prediction. (Figure from Ref. \cite{zg_prd}, see text)}
\label{unfolded}
\end{figure}

\begin{figure}[htbp]
\centering
\includegraphics[width=0.45\textwidth]{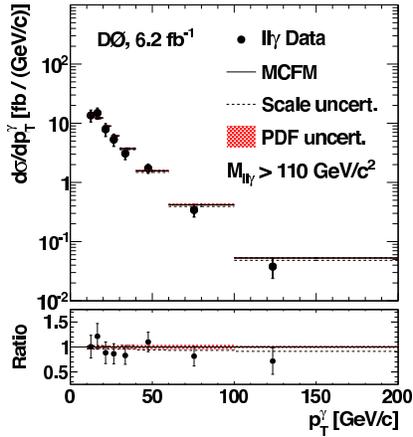}
\caption{Unfolded $d\sigma/dp_T^\gamma$ distribution with $M_{ll\gamma}>110$ GeV/$c^2$ for combined electron and muon data compared with the NLO {\sc mcfm} prediction. (Figure from Ref. \cite{zg_prd}, see text)}
\label{unfolded_llg110}
\end{figure}

The matrix inversion technique \cite{unfolding} is used to extract the
differential cross section $d\sigma/dp_{T}^{\gamma}$ as a function of
the true $p_{T}^{\gamma}$.
The {\sc pythia} $Z\gamma$ events are used to assemble the smearing matrix
between true and reconstructed $p_{T}^{\gamma}$.
In the end, the matrix is inverted to obtain the unsmeared spectrum.
The measured differential cross sections $d\sigma/dp_T^\gamma$ are shown in
Figs.~\ref{unfolded} and \ref{unfolded_llg110} for no $M_{ll\gamma}$
requirement and $M_{ll\gamma}>110$ GeV, respectively.
The values associated with Figs.~\ref{unfolded} and \ref{unfolded_llg110}
are given in Tables \ref{unfoldedvals} and \ref{unfoldedvals_llg110}.
The measured cross section are good consistent with NLO {\sc mcfm}.

\begin{table*}
\begin{center}
\caption{\label{unfoldedvals} Summary of the unfolded differential cross section $d\sigma/dp_T^\gamma$, when no $M_{\ell\ell\gamma}$ requirement is applied, and NLO {\sc mcfm} predictions with PDF and scale uncertainties.
(Table from Ref. \cite{zg_prd}, see text)}
  \begin{tabular}{ D{,}{-}{-1}   c  l  l  }
  \hline \hline
     \multicolumn{2}{c}{ }  & \multicolumn{1}{c}{$\ell\ell\gamma$ Combined Data} & \multicolumn{1}{c}{NLO {\sc mcfm}} \\ \hline
   \multicolumn{1}{c}{$p_T^\gamma$ bin }  & $p_T^\gamma$ center & \multicolumn{2}{c}{$d\sigma/d  p_T^\gamma $ }    \\
    \multicolumn{1}{c}{\mbox{[GeV/$c$]}} &  [GeV/$c$] & \multicolumn{2}{c}{ [fb/(GeV/$c$)]}     \\ \hline
   10,15 & 12.4         &       111.14 $\pm$ 4.40 (stat.)~$\pm$ 11.99 (syst.)   & 104.02 $\pm$ 4.10 (PDF)   $^{+1.4}_{-1.2}$ (scale)            \\
   15 , 20 & 17.2       &       51.41 $\pm$ 3.83 (stat.)~$\pm$ 2.65 (syst.)             &  57.13 $\pm$ 2.23 (PDF)  $^{+1.3}_{-1.8}$ (scale)             \\
   20 , 25 & 22.5       &       25.34 $\pm$ 2.74 (stat.)~$\pm$ 1.13      (syst.)        & 28.77 $\pm$ 0.43 (PDF)  $^{+1.1}_{-0.7}$ (scale)              \\
   25 , 30 & 27.5       &       8.08 $\pm$ 1.45 (stat.)~$\pm$ 0.40      (syst.)         & 10.16 $\pm$ 0.26 (PDF)  $^{+0.7}_{-0.5}$ (scale)              \\
   30 , 40 & 34.4       &       3.23 $\pm$ 0.60 (stat.)~$\pm$ 0.17      (syst.)         & 4.15 $\pm$ 0.16 (PDF)    $^{+0.34}_{-0.19}$ (scale)                   \\
   40 , 60 & 48.5       &       1.70 $\pm$ 0.26 (stat.)~$\pm$ 0.088      (syst.)        & 1.60 $\pm$ 0.061 (PDF)  $^{+0.008}_{-0.010}$ (scale)                  \\
   60 , 100 & 76.5      &       0.34 $\pm$ 0.079 (stat.)~$\pm$ 0.018 (syst.)    & 0.42 $\pm$ 0.017 (PDF)    $^{+0.028}_{-0.028}$ (scale)                \\
   100 , 200 & 124.5    &       0.038 $\pm$ 0.014 (stat.)~$\pm$ 0.002 (syst.)   & 0.052 $\pm$ 0.001 (PDF)    $^{+0.003}_{-0.001}$ (scale)               \\
   \hline \hline
  \end{tabular}
\end{center}
\end{table*}

\begin{table*}
\begin{center}
\caption{\label{unfoldedvals_llg110} Summary of the unfolded differential cross section $d\sigma/dp_T^\gamma$, with the $M_{\ell\ell\gamma}>110$ GeV/$c^2$ requirement, and NLO {\sc mcfm} predictions with PDF and scale uncertainties.
(Table from Ref. \cite{zg_prd}, see text)}
 \begin{tabular}{ D{,}{-}{-1}   c  l   l  }
   \hline \hline
     \multicolumn{2}{c}{ }  & \multicolumn{1}{c}{$\ell\ell\gamma$ Combined Data} & \multicolumn{1}{c}{NLO {\sc mcfm}}\\ \hline
   \multicolumn{1}{c}{$p_T^\gamma$ bin }  & $p_T^\gamma$ center & \multicolumn{2}{c}{$d\sigma/d  p_T^\gamma $ }    \\
    \multicolumn{1}{c}{\mbox{[GeV/$c$]}} &  [GeV/$c$] & \multicolumn{2}{c}{ [fb/(GeV/$c$)]}     \\ \hline
   10 , 15 & 13.7       &       13.57 $\pm$ 1.87 (stat.)~$\pm$ 2.43 (syst.)             & 13.48 $\pm$ 0.48 (PDF) $^{+0.25}_{-0.51}$ (scale)             \\
   15 , 20 & 17.2       &       14.87 $\pm$ 2.17 (stat.)~$\pm$ 2.30 (syst.)             & 12.25 $\pm$ 0.47 (PDF)  $^{+0.29}_{-0.36}$ (scale)                    \\
   20 , 25 & 22.0       &       7.91 $\pm$ 1.76 (stat.)~$\pm$ 0.81      (syst.)         & 8.94 $\pm$ 0.25 (PDF) $^{+0.13}_{-0.35}$ (scale)              \\
   25 , 30 & 27.4       &       5.30 $\pm$ 1.15 (stat.)~$\pm$ 0.44      (syst.)         & 6.13  $\pm$ 0.21(PDF) $^{+0.016}_{-0.25}$ (scale)     \\
   30 , 40 & 34.5       &       3.08 $\pm$ 0.57 (stat.)~$\pm$ 0.33      (syst.)         & 3.71  $\pm$ 0.15  (PDF) $^{+0.012}_{-0.14}$ (scale)                   \\
   40 , 60 & 48.6       &       1.73 $\pm$ 0.26 (stat.)~$\pm$ 0.17      (syst.)         & 1.57 $\pm$ 0.061 (PDF) $^{+0.004}_{-0.094}$ (scale)           \\
   60 , 100 & 76.5      &       0.34 $\pm$ 0.079 (stat.)~$\pm$ 0.019 (syst.)            & 0.42 $\pm$ 0.017 (PDF) $^{+0.028}_{-0.028}$ (scale)                   \\
   100 , 200 & 124.5    &       0.038 $\pm$ 0.014 (stat.)~$\pm$ 0.002 (syst.)   & 0.052 $\pm$ 0.001  (PDF) $^{+0.003}_{-0.001}$ (scale)                 \\
   \hline \hline
  \end{tabular}
\end{center}
\end{table*}

\clearpage
\section{Limits on anomalous $WW\gamma$, $ZZ\gamma$ and $Z\gamma\gamma$ couplings}
For $W\gamma$ production, an effective Lagrangian parameterizes the
$WW\gamma$ couplings with two parameters,
$\kappa_{\gamma}$ and $\lambda_{\gamma}$ \cite{baur-ref1, baur-ref2, baur-ref3},
under the assumptions of electromagnetic gauge invariance,
charge conjugation ($C$), parity ($P$) and $CP$ conservation.
The $\kappa_{\gamma}$ and $\lambda_{\gamma}$ couplings are related to the
magnetic dipole and electric quadrupole moments
of the $W$ boson.
In the SM, $\kappa_{\gamma} = 1$ and $\lambda_{\gamma} = 0$,
and it is customary to introduce into the notation
the difference $\Delta\kappa_{\gamma} \equiv \kappa_{\gamma} - 1$.
A form factor with a 2 TeV common scale $\Lambda$ for each
non-SM coupling parameter, is used to assure that the $W\gamma$ cross section
does not violate unitarity.

\begin{figure}[htbp]
\centering
\subfigure[]{\label{fig:ISR1}\includegraphics[width=0.35\textwidth]{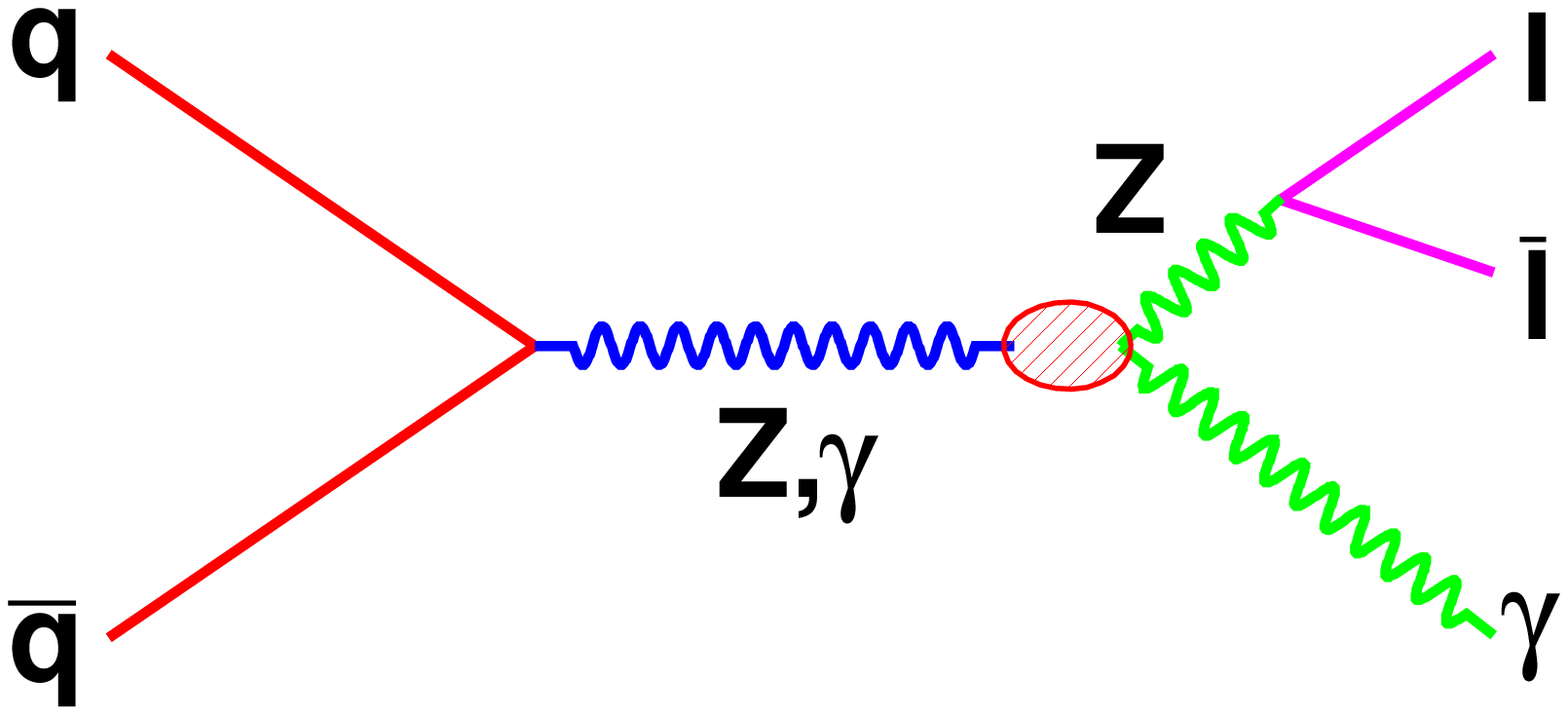}}
\subfigure[]{\label{fig:ISR2}\includegraphics[width=0.35\textwidth]{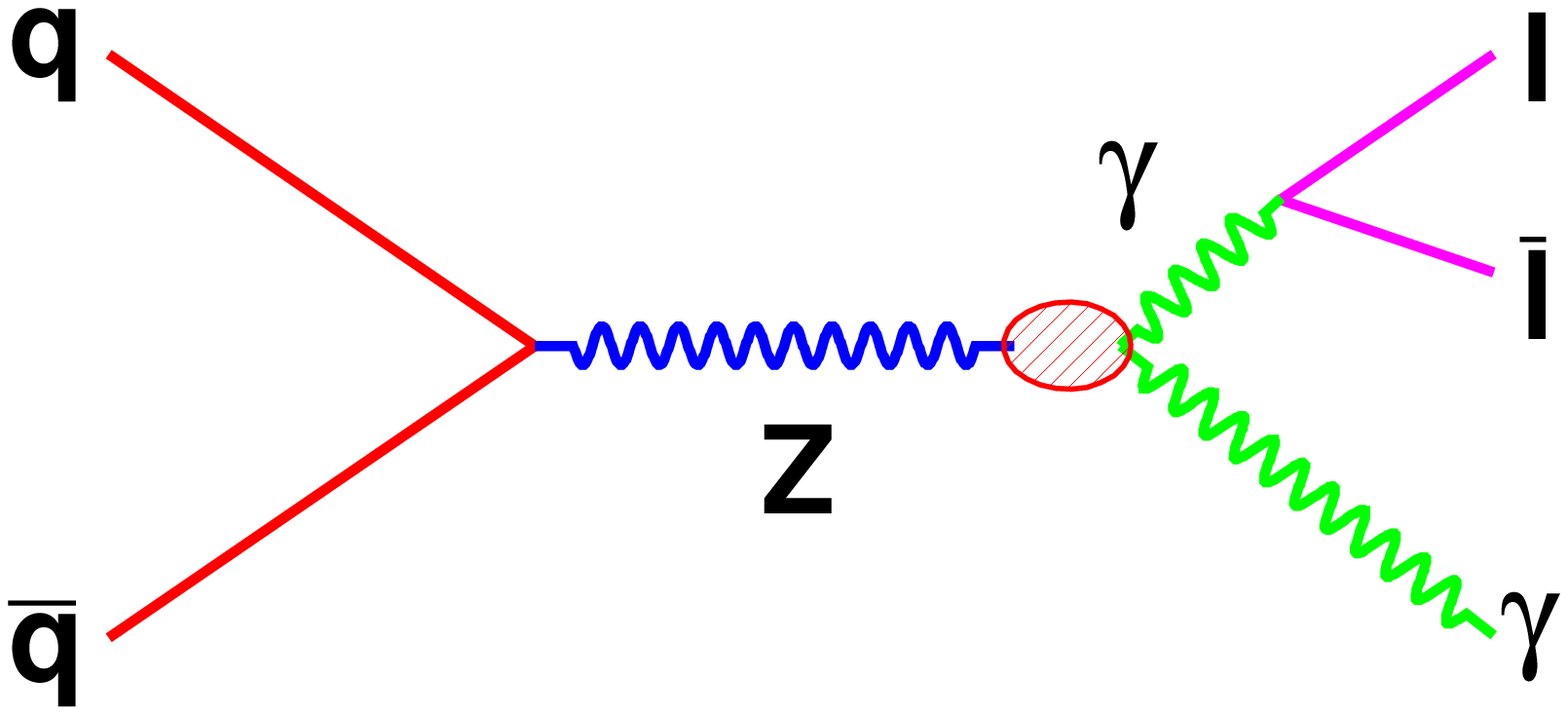}}
\caption{\small
  Feynman diagrams for the anomalous $ZZ\gamma$ and $Z\gamma\gamma$ couplings.
 }
\label{fig:ZG_AC}
\end{figure}

For the $Z\gamma$ production, an effective theory with
eight complex coupling parameters, $h^V_i$, where $i=1,2,3,4$ and $V=Z$
or $\gamma$ \cite{hzp} is introduced to describe the anomalous $ZZ\gamma$
and $Z\gamma\gamma$ couplings (see Fig. \ref{fig:ZG_AC}).
A form factor with a common scale $\Lambda$ is introduced to
conserve tree-level unitarity at
asymptotically high energies.
The $\Lambda = 1.2$ and $1.5$ TeV are used to set the limits on
$CP$-even coupling parameters $h^V_{03}$ and $h^V_{04}$.

The contributions from anomalous couplings will increase the $W\gamma$
and $Z\gamma$ production cross section, and particularly give
rise to an excess of photons in high transverse momentum.
Thus, the photon transverse momentum distribution is used as the golden
candle to search for anomalous departure from SM $WW\gamma$, $ZZ\gamma$
and $Z\gamma\gamma$ couplings.

\begin{figure}[htbp]
 \centering
\hspace*{-9mm} \includegraphics[scale=0.5]{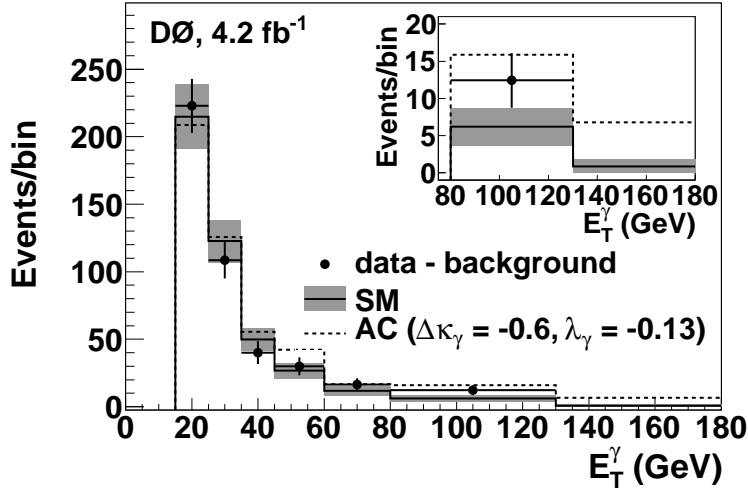}
~\\[-4mm]
   \caption{\small Photon transverse energy distributions
for background-subtracted data compared to the expectation for the SM
and for one choice of anomalous couplings
for the combined electron and muon channels.
The background-subtracted data are shown as black points
with uncertainties representing the associated
statistical and systematic uncertainties.
The SM prediction is given by the solid line,
with the shaded area representing its uncertainty.
The effect of one example of anomalous couplings
is represented by the dashed line.
The last $p_T^{\gamma}$ bin shows the sum
of all events with $p_T^{\gamma} > 130$ GeV.
The inset shows the distributions in the last two bins of $p_T^{\gamma}$.
(Figure from Ref. \cite{wg_prl}, see text)
}
  \label{figure-gpt}
\end{figure}

\begin{figure}[htbp]
 \centering
\hspace*{-9mm} \includegraphics[scale=0.41]{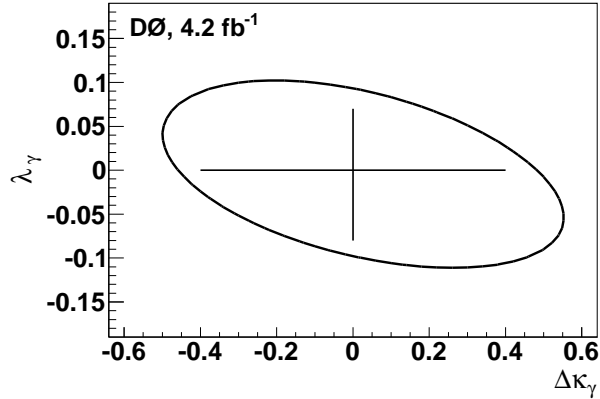}
~\\[-4mm]
   \caption{\small
  Limits on the $WW\gamma$ coupling parameters $\Delta\kappa_{\gamma}$
and $\lambda_{\gamma}$. The ellipse represents the two-dimensional 95\% C.L.
exclusion contour. The one-dimensional 95\% C.L. limits are shown as
the vertical and horizontal lines. (Figure from Ref. \cite{wg_prl}, see text)
}
  \label{figure-AClimits}
\end{figure}

The photon $p_{T}^{\gamma}$ distributions from $W\gamma$ production
is shown in Fig.~\ref{figure-gpt},
and used to derive limits on anomalous $WW\gamma$ trilinear couplings
using a binned likelihood fit to data.
The 95\%~C.L. limits on the $WW\gamma$ coupling parameters
$\Delta\kappa_{\gamma}$ and $\lambda_{\gamma}$ are shown
in Fig.~\ref{figure-AClimits}, with the contour defining the two-dimensional
exclusion limits.
The one-dimensional 95\%~C.L. limits are
$-0.4 < \Delta\kappa_{\gamma} < 0.4$
and $-0.08 < \lambda_{\gamma} < 0.07$,
which are obtained by setting one coupling parameter to the SM value
and allowing the other to vary.
These are the most stringent limits on anomalous $WW\gamma$ couplings
at a hadron collider.

\begin{figure}[h!]
\begin{minipage}[b]{0.9\linewidth}
\centering
\includegraphics[scale=0.45]{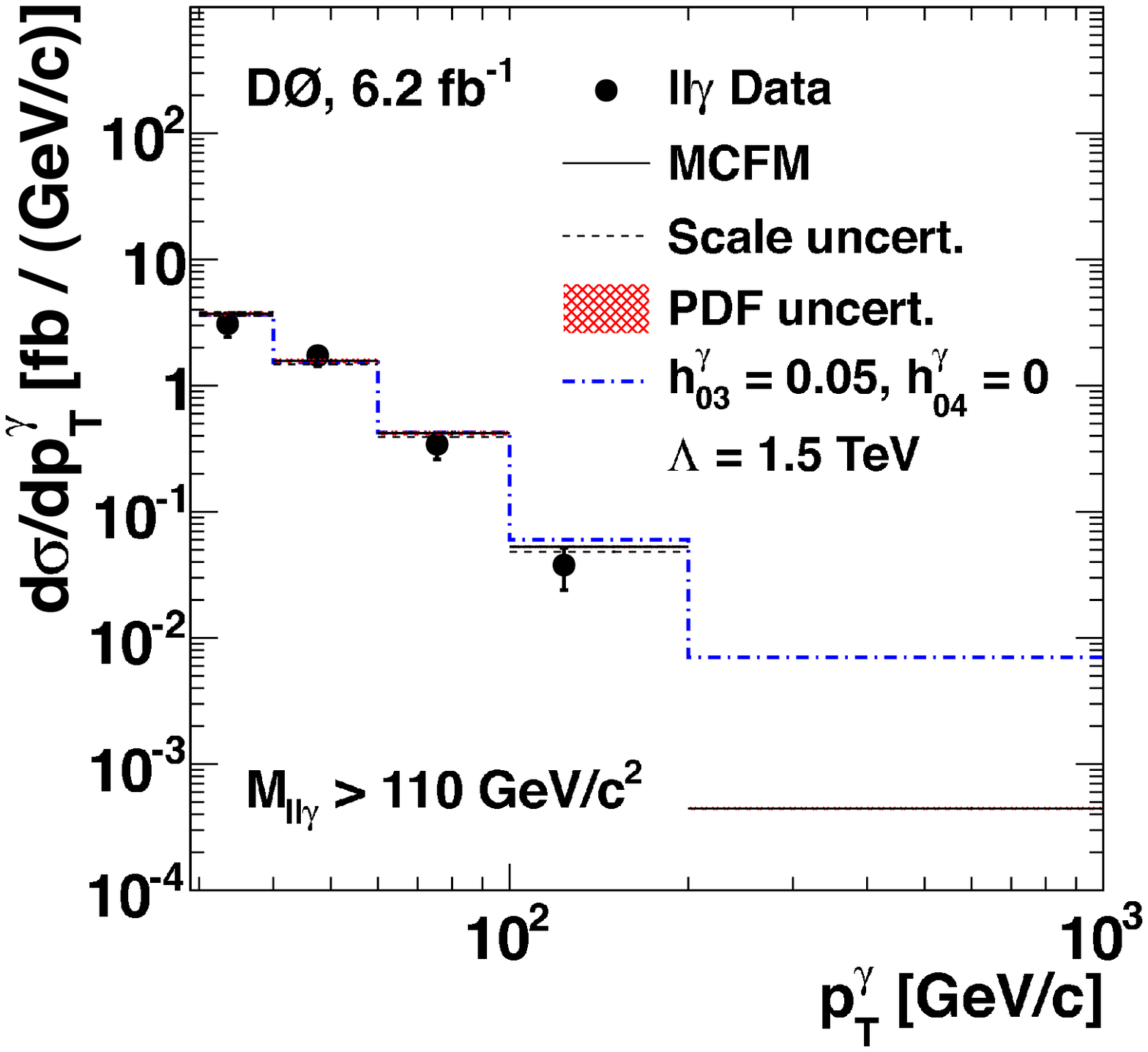}
\caption{\label{ACplot} The SM prediction and anomalous $Z\gamma$ coupling production compared with the unfolded $d\sigma/dp_T^\gamma$ for combined muon and electron channels for $p_T^\gamma > 30$ GeV and $M_{ll\gamma}>110$ GeV.
(Figure from Ref. \cite{zg_prd}, see text)}
\end{minipage}
\end{figure}

\begin{figure}
\subfigure[]{\label{fig:ISR1}\includegraphics[scale=.4]{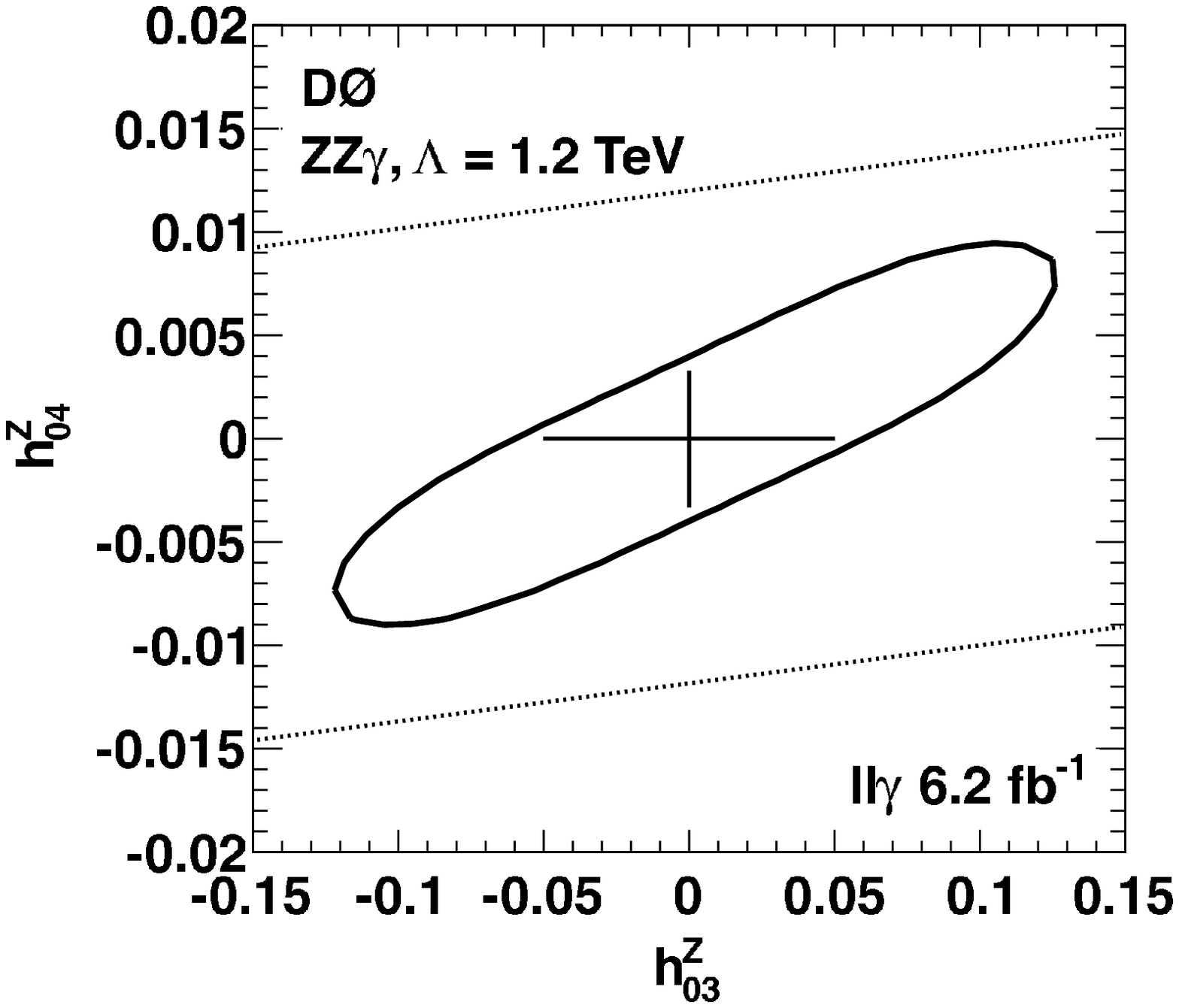}}
\subfigure[]{\label{fig:ISR1}\includegraphics[scale=.4]{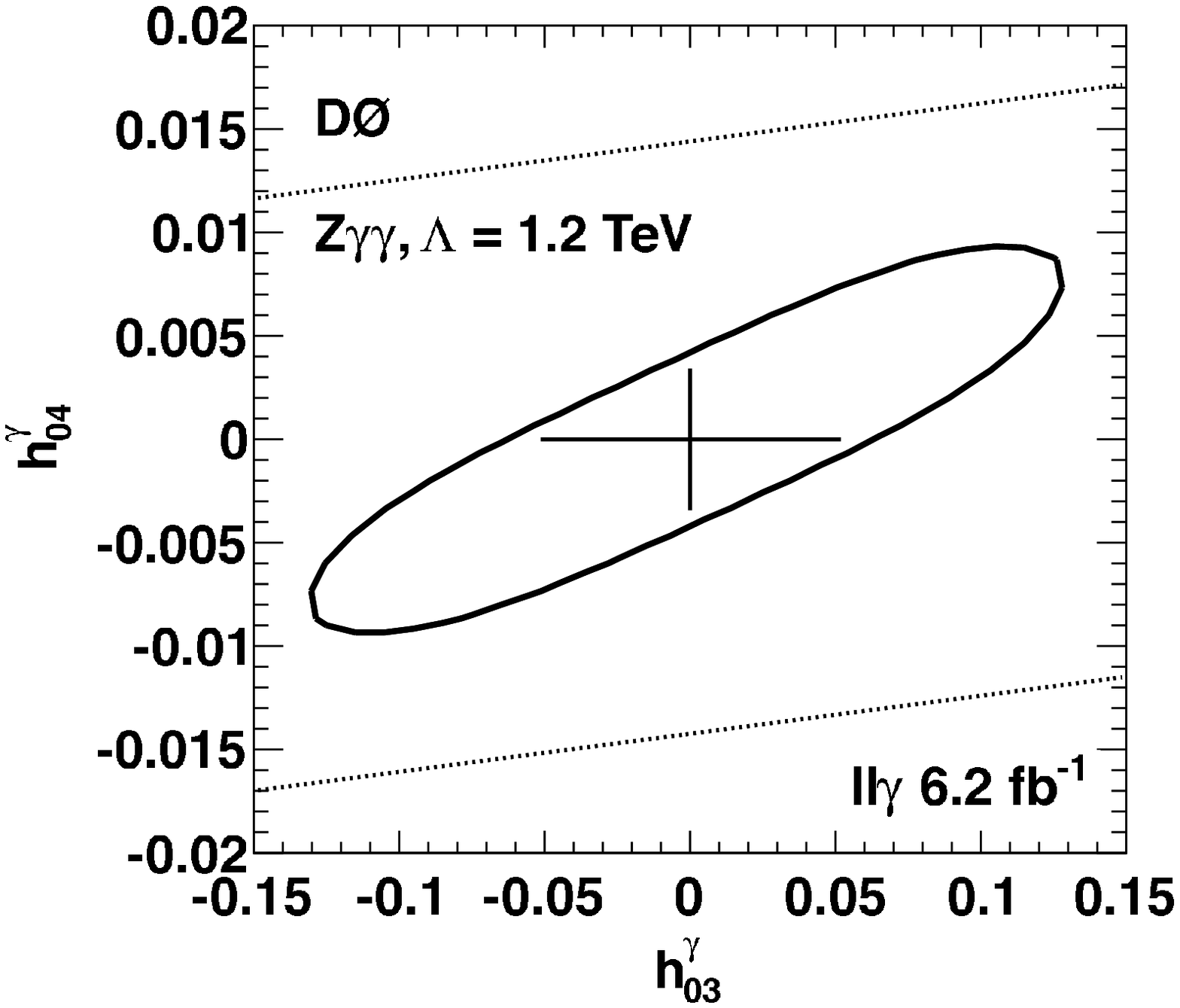}}
\caption{The 2D (contour) and 1D (cross) limits on the anomalous coupling parameters for (a) $ZZ\gamma$ and (b)  $Z\gamma\gamma$ vertices at the 95\% C.L. for $\Lambda=1.2$ TeV.  Limits on $S$-matrix unitarity are represented by the dotted lines. (Figure from Ref. \cite{zg_prd}, see text)}
\label{2D_1200}
\end{figure}

\begin{figure}
\subfigure[]{\label{fig:ISR1}\includegraphics[scale=.4]{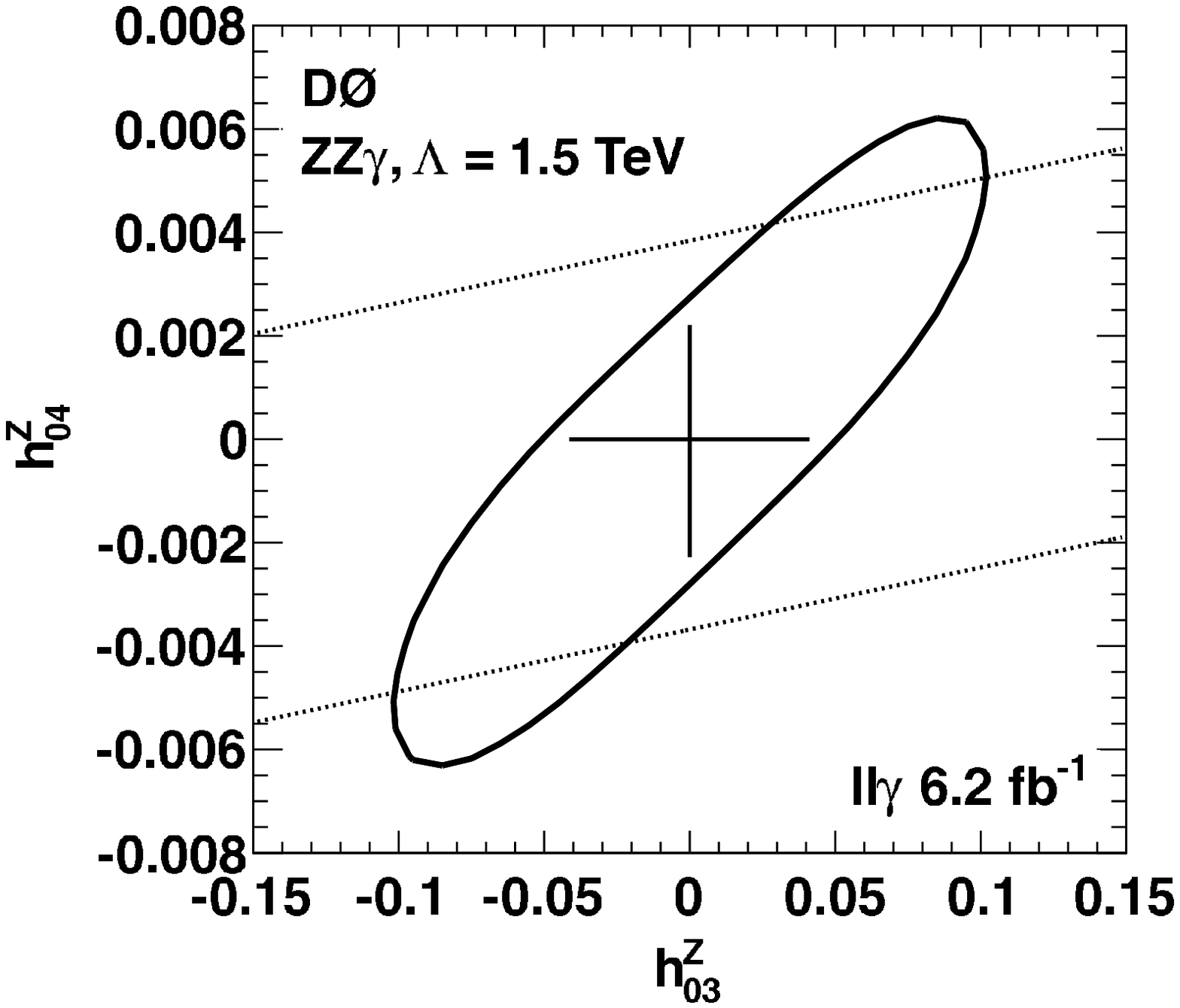}}
\subfigure[]{\label{fig:ISR1}\includegraphics[scale=.4]{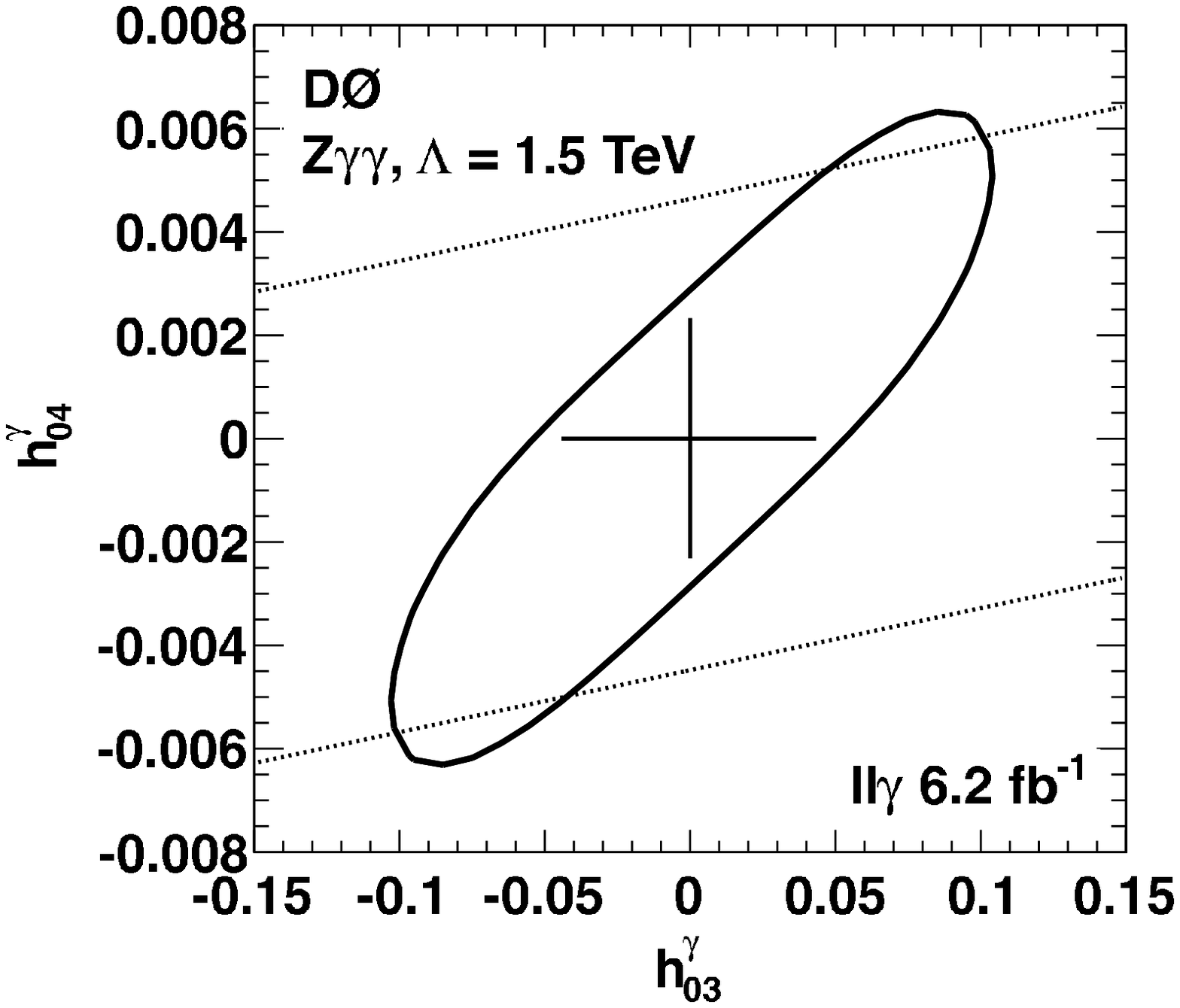}}
\caption{The 2D (contour) and 1D (cross) limits on the anomalous parameters for (a) $ZZ\gamma$ and (b)  $Z\gamma\gamma$ vertices at the 95\% C.L. for $\Lambda=1.5$ TeV.  Limits on $S$-matrix unitarity are represented by the dotted lines.
(Figure from Ref. \cite{zg_prd}, see text)}
\label{2D_1500}
\end{figure}

\begin{figure}[h!]
\begin{minipage}[b]{1.0\linewidth}
\subfigure[]{\label{fig:ISR1}\includegraphics[scale=.45]{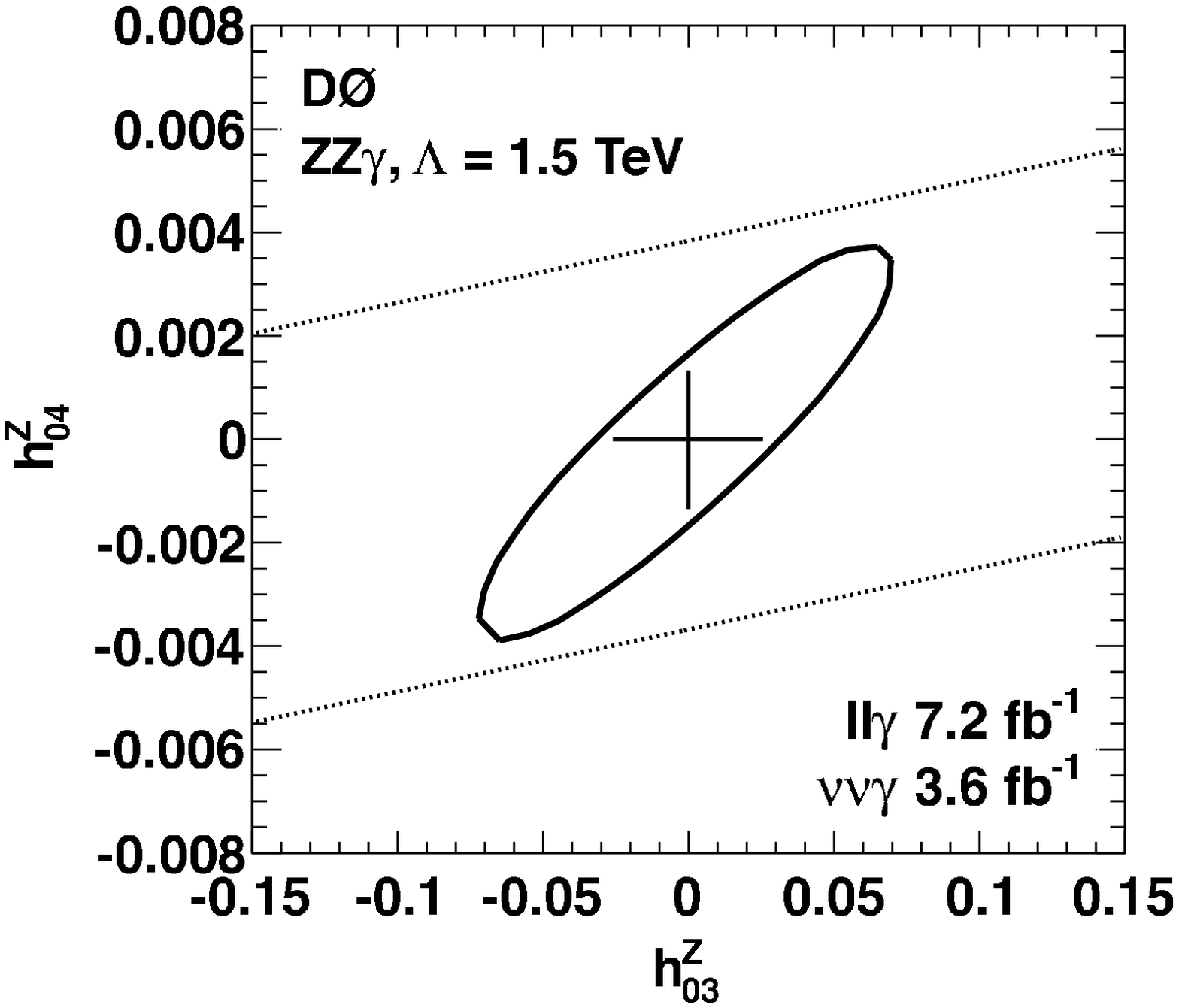}}
\subfigure[]{\label{fig:ISR1}\includegraphics[scale=.45]{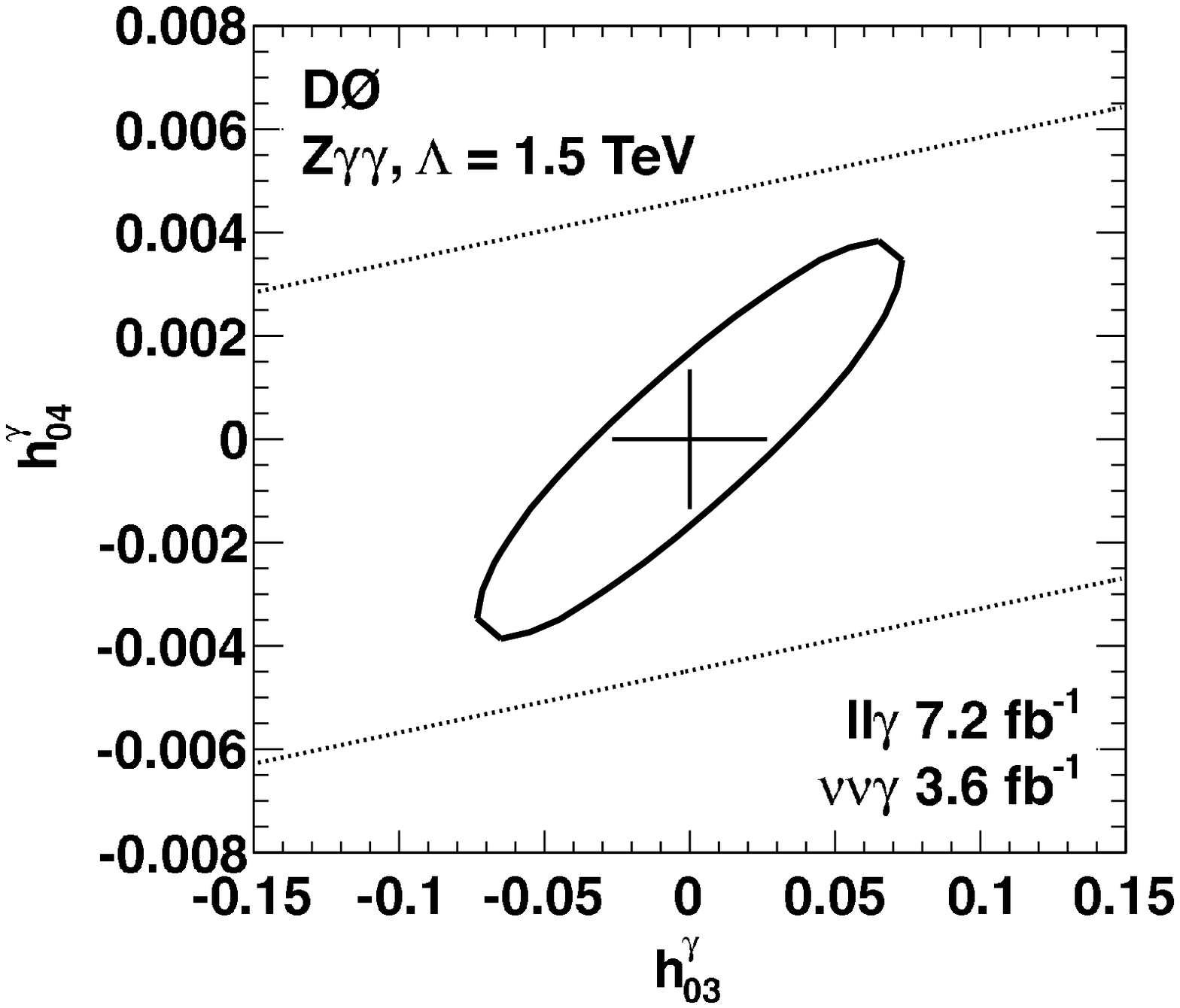}}
\caption{\label{2D_1500_wprev} 2D (contour) and 1D (cross) limits on coupling parameters for (a) $ZZ\gamma$ and (b)  $Z\gamma\gamma$ vertices at the 95\% C.L. for $\Lambda=1.5$ TeV.  Limits on $S$-matrix unitarity are represented by the dotted lines.
(Figure from Ref. \cite{zg_prd}, see text)}
\end{minipage}
\end{figure}

The $d\sigma/dp_T^{\gamma}$ distributions of $Z\gamma$ production is
shown in Fig. \ref{ACplot}, and folded into a reconstruction-level
distribution to derive the limits on anomalous coupling parameters
$h^V_{03}$ and $h^V_{04}$.
To eliminate the FSR contribution, only events with $p_T^{\gamma} > 30$ GeV
and $M_{ll\gamma} > 110$ GeV are considered.
The 1D limits for $\Lambda = 1.2$ TeV and 1.5 TeV are shown in
Table \ref{1Dlimits}, and the corresponding plots are shown in
Figs. \ref{2D_1200} and \ref{2D_1500}.
After combining the previous 1 fb$^{-1}$ D0 $Z\gamma$ analysis \cite{d0prev},
the limits are shown in Fig.~\ref{2D_1500_wprev} and Table \ref{1Dlimits}.

\begin{table}[h!]
\caption{\label{1Dlimits} Summary of the 1D limits on the $ZZ\gamma$ and $Z\gamma\gamma$ coupling parameters at the 95\% C.L.
(Table from Ref. \cite{zg_prd}, see text) }
\begin{center}
  \begin{tabular}{   l  c  c  c }
\hline \hline
  & &  &  $ll\gamma$ 7.2 fb$^{-1}$ \\
  & \multicolumn{2}{c}{$ll\gamma$ 6.2 fb$^{-1}$} & $\nu\nu\gamma$ 3.6 fb$^{-1}$ \\
  & $\Lambda=1.2$ TeV & $\Lambda=1.5$ TeV & $\Lambda = 1.5$ TeV \\ \hline
  $|h^Z_{03}|<$ & 0.050 & 0.041 & 0.026\\
  $|h^Z_{04}|<$ & 0.0033 & 0.0023 & 0.0013\\ \hline
   $|h^\gamma_{03}|<$ & 0.052 & 0.044 & 0.027\\
  $|h^\gamma_{04}|<$ & 0.0034 & 0.0023 & 0.0014\\ \hline \hline
\end{tabular}
\end{center}
\end{table}

\section{Conclusions and outlook}
D0 has performed sophisticated studies of $W\gamma$ and $Z\gamma$ 
production using about half of the collected dataset.
The results include the most precise measurements of the total production
cross section of $W\gamma \to l\nu\gamma$ and $Z\gamma \to ll\gamma$,
and the first unfolded photon differential cross section 
$d\sigma/dp_T^{\gamma}$, as well as the most stringent limits
on the $WW\gamma$ anomalous couplings. The D0 dataset will be continuously 
analyzed, and the high energy LHC experiments have already taken 
and analyzed data. 
The prospects for anomalous triple gauge boson couplings will be interesting
and fruitful.


\begin{thebibliography}{50}
\bibitem{d0_coordinate}
Pseudorapidity is defined as $\eta=-\ln [\tan(\theta/2)]$, where
$\theta$ is the polar angle relative to the proton beam direction.
$\phi$ is defined to be the azimuthal angle in the plane transverse to
the proton beam direction.
\end{thebibliography}
\end{document}